\documentclass[notitlepage,aps,longbibliography,letter,reprint,footinbib]{revtex4-1}

\usepackage{graphicx}
\usepackage{amsmath}
\usepackage{amssymb}
\usepackage{hyperref}
\usepackage{color}
\usepackage[abs]{overpic}
\usepackage[english]{babel}

\usepackage{letltxmacro}

\LetLtxMacro{\ORIGselectlanguage}{\selectlanguage}
\makeatletter
\DeclareRobustCommand{\selectlanguage}[1]{%
  \@ifundefined{alias@\string#1}
    {\ORIGselectlanguage{#1}}
    {\begingroup\edef\x{\endgroup
       \noexpand\ORIGselectlanguage{\@nameuse{alias@#1}}}\x}%
}
\newcommand{\definelanguagealias}[2]{%
  \@namedef{alias@#1}{#2}%
}
\makeatother

\definelanguagealias{en}{english}

\newcommand{\be}{\begin{eqnarray*}}
\newcommand{\ee}{\end{eqnarray*}}
\newcommand{\beq}{\begin{eqnarray}}
\newcommand{\eeq}{\end{eqnarray}}
\newcommand{\bequ}{\begin{equation}}
\newcommand{\eequ}{\end{equation}}
\newcommand{\bk}{{\mathbf{k}}}

\newcommand{\bp}{{\mathbf{p}}}

\newcommand{\br}{{\mathbf{r}}}

\newcommand{\bA}{{\mathbf{A}}}

\newcommand{\dd}{\mathrm{d}}

\newcommand{\bx}{\hat{{\mathbf{x}}}}
\newcommand{\by}{\hat{{\mathbf{y}}}}
\newcommand{\h}{\hat{H}}

\newcommand{\ph}{{\phantom{\dagger}}}
\newcommand{\ket}[1]{\left|{#1}\right\rangle}
\newcommand{\bra}[1]{\left\langle{#1}\right|}
\newcommand{\hc}{\mathrm{H.c.}}
\newcommand{\an}[1]{\left(a\right)^{#1}}
\newcommand{\ad}[1]{\left(a^\dagger\right)^{#1}}
\newcommand{\ado}{a^\dagger}

\newcommand{\kx}{\hat{k}_x}
\newcommand{\ky}{\hat{k}_y}

\newcommand{\id}{\mathbb{I}}
\newcommand{\mcs}{\mathcal{S}}
\newcommand{\ta}{\mathtt{a}}

\begin{document}
\title{Finite-wavevector Electromagnetic Response in Lattice Quantum Hall Systems}
\author{Fenner Harper}
\author{David Bauer}
\author{T.~S.~Jackson}
\author{Rahul Roy}
\affiliation{Department of Physics and Astronomy, University of California, Los Angeles, California USA}
\date{\today}
\begin{abstract}
In a quantum Hall system, the finite-wavevector Hall conductivity displays an intriguing dependence on the Hall viscosity, a coefficient that describes the non-dissipative response of the fluid to a velocity gradient. In this paper, we pursue this connection in detail for quantum Hall systems on a lattice, noting that the neat continuum relation breaks down and develops corrections due to the broken rotational symmetry. In the process, we introduce a new, quantum mechanical derivation of the finite-wavevector Hall conductivity for the integer quantum Hall effect, which allows terms to arbitrary order in the wavevector expansion to be calculated straightforwardly. We also develop a universal formalism for studying quantum Hall physics on a lattice, and find that at weak applied magnetic fields, generic lattice wavefunctions connect smoothly to the Landau levels of the continuum. At moderate field strengths, the lattice corrections can be significant and perturb the wavefunctions, energy levels, and transport properties from their continuum values. Our approach allows the finite-field behaviour of a system to be inferred directly from the zero-field band structure.

\end{abstract}
\maketitle
\section{Introduction}
The quantum Hall effects \cite{prange1987quantum} are impressive examples of macroscopic quantum phenomena with measurable experimental signatures \cite{Klitzing:428079,Tsui:1982yy}. Distinctively, quantum Hall phases exhibit a precisely quantised transverse (Hall) conductance, $\sigma_{xy}$, which derives from the topological properties of the underlying wavefunction. In an experiment, this is manifested as a characteristic sequence of resistance plateaux that arise as the strength of an external magnetic field is swept. In addition to this, quantum Hall systems also exhibit a precisely quantised Hall viscosity, $\eta_H$, a transport coefficient that describes the non-dissipative response of the quantum Hall fluid to a velocity gradient \cite{Lifshitz:2012th,Levay:1995dq,Avron:276305}. Like $\sigma_{xy}$, the Hall viscosity also derives from the topology of the underlying wavefunction \cite{Read:2009iw,Read:2011hx}. It may be expressed more fundamentally in terms of the shift or mean orbital spin \cite{Wen:1992ug,Read:2009iw,Read:2011hx}, quantised topological numbers, unique to a particular quantum Hall phase, that describe the behaviour of the fluid in a system with curvature. While difficult to measure directly, it has recently been shown that the Hall viscosity occurs as a coefficient in the finite-momentum expansion of the Hall conductivity, $\sigma_{xy}(q)$, in the presence of an inhomogeneous electric field \cite{Hoyos:2012eu,Bradlyn:2012he,Biswas:2013wm,Haldane:2015tx}. This paves the way for an experimental measurement of Hall viscosity \cite{Huang:2015ga,Scaffidi:2017fb,Pellegrino:2017fd,Delacretaz:2017kx,Gromov:2017io} and, consequently, a new diagnostic tool for identifying quantum Hall phases. The intriguing connection between conductivity and viscosity has been demonstrated using field theory \cite{Hoyos:2012eu}, Kubo formulae \cite{Bradlyn:2012he}, and semiclassical arguments \cite{Biswas:2013wm}, and in this article we complement these approaches with a quantum mechanical treatment. 

Most studies of the QHE, however, make simplifying assumptions about the host medium: namely, that there is continuous translational and rotational symmetry, and that the underlying system is well described by the two-dimensional electron gas (2DEG). In reality, quantum Hall experiments take place in semiconductor heterostructures \cite{Klitzing:428079,Tsui:1982yy,Contreras:1997uc,Tsukazaki:2007be}, or in more exotic systems such as optical lattices of cold atoms \cite{Miyake:2013jw,Aidelsburger:2013ew}, in suspended layers of graphene \cite{Zhang:2005gp,Novoselov:2005es}, and on the surfaces of 3D topological insulators \cite{Xu:2014eh}. In most of these settings, the system is not continuous, and instead has a periodic lattice structure. In general, the lattice endows the system with a band structure that can have significant effects on the physics. Notably, the single-particle wavefunctions tend to form dispersive energy bands, rather than perfectly flat Landau levels, while the quantum geometry of the problem, codified in the Berry curvature and quantum metric \cite{Berry:1984ka,Parameswaran:2012uk,Roy:2014dl}, can significantly alter the wavefunction properties \cite{Bergholtz:2013ey,Parameswaran:2013pca,Harper:2014fq,Bauer:2016ju}. In turn, these single-particle effects alter the transport properties and stability of many-body, fractional quantum Hall phases \cite{Xiao:2010kw,Jackson:2015fv}. When these effects are strong, a lattice quantum Hall system may more accurately be described as a Chern insulator (see Refs.~\cite{Bergholtz:2013ey,Parameswaran:2013pca} for reviews). In these systems, the lattice substantially alters the physics of the continuum, but may also support quantum Hall-like states in the absence of a net external magnetic field \cite{Haldane:1988uf,Neupert:2011db,Sun:2011dk,Sheng:2011iv,Chang:2013dd}, raising the possibility of realising the phenomenon at room temperature \cite{He:2017ib}.

In many realisations of the quantum Hall effect, the magnetic flux per lattice plaquette is small and the continuum QHE gives a good approximation of the behaviour of the system. Heuristically, this is because the characteristic length scale of the wavefunctions, which scales with the magnetic flux density as $1/\sqrt{B}$, is much larger than the characteristic length scale of the lattice, and so the discreteness of the lattice is not detected. However, the realisation of stronger magnetic fields, the advent of techniques to simulate gauge fields in systems of cold atoms \cite{Dalibard:2011gg}, and the discovery of quantum Hall physics in graphene superlattices \cite{Ponomarenko:2014hl,Dean:2014bv}, have brought the strong-field regime into sharp focus. In this limit, lattice corrections have significant, measurable consequences, which would need to be accounted for in a putative experiment to measure the Hall viscosity.

In this article, we consider such lattice effects in detail. As inspiration, we take the well-studied Hofstadter model \cite{Harper:1955uu,Hofstadter:1976wt} (see Ref.~\onlinecite{Harper:2014fq} for a review), a simple tight binding model for the QHE on a lattice that allows interpolation between strong- and weak-field regimes. As we show, generic tight-binding models with an external magnetic field support Hofstadter-like eigenstates, and in the weak field limit connect smoothly to continuum states. We use perturbative methods to derive the finite-wavevector transverse electromagnetic response as a series expansion in the flux density. In the process, we generalise a recent perturbative study of the Hofstadter model \cite{Harper:2014fq,Bauer:2016ju} to arbitrary lattices and simplify the approach through an extension of the lattice Peierls substitution. In many cases, lattice corrections to the wavefunction and transport coefficients may be obtained simply through knowledge of the relevant band structure, a quantity that may be readily obtained in experiments or numerics.

The paper is structured as follows. In Sec.~\ref{sec:em_continuum}, we give a quantum mechanical derivation of the finite-wavevector Hall conductivity, $\sigma_{xy}(q)$. This is motivated by, but distinct from, a recent semiclassical derivation of this quantity given in Ref.~\cite{Biswas:2013wm}. In Sec.~\ref{sec:perturbationtheory}, we outline a general perturbative approach for obtaining lattice corrections to wavefunctions and energy levels when an underlying tight-binding model or dispersion relation is given. We go on, in Sec.~\ref{sec:em_lattice}, to use these perturbed quantities to obtain the lattice corrections to $\sigma_{xy}(q)$. General lattices are considered, with specific results given for those with $C_4$ symmetry and for the original Hofstadter model. In Sec.~\ref{sec:conclusions} we discuss our results and conclude. Some of the more mathematical parts of the paper may be found in the Appendices.

\section{Electromagnetic Response in the Continuum\label{sec:em_continuum}}
\subsection{Preliminary Discussion}
In this section, we give a new, quantum mechanical derivation of the finite-wavevector electromagnetic response of an integer quantum Hall fluid. This approach will then be used in more complex settings in Sec.~\ref{sec:em_lattice}, where we calculate the electromagnetic response of lattice quantum Hall systems.

We first recall some well-established results from the literature. In a quantum Hall system, an applied electric field in the $x$-direction generates a transverse current density according to the relation
\beq
J_y(q)&=&\sigma_{xy}(q)E_x(q),\label{eq:j_sigma_e}
\eeq
where $\sigma_{xy}(q)$ is the Hall conductivity. In the limit $q\to0$, the Hall conductivity takes a universal quantised value $\sigma_{xy}(0)=\nu e^2/(2\pi\hbar)$, where $\nu$ is a rational number that gives the filling fraction of the Landau level. If the electric field is inhomogeneous, the Hall conductivity is dependent on the wavevector $q$. Several references \cite{Hoyos:2012eu,Bradlyn:2012he,Biswas:2013wm,Haldane:2015tx} have shown that the leading term in a series expansion in $q$ arises at $O(q^2)$ and has a coefficient that depends on the Hall viscosity $\eta_H$. Combining results from Ref.~\cite{Hoyos:2012eu}, $\sigma_{xy}(q)$ may be expanded as
\beq
\sigma_{xy}(q)&=&\sigma_{xy}(0)\bigg[1+\left(ql_B\right)^2\left(\frac{\eta_H}{\hbar\rho_0}-\frac{1}{\nu}\frac{2\pi l_B^2}{\hbar\omega_c}B^2\epsilon''(B)\right)\nonumber\\
&&+O\left((ql_B)^4\right)\bigg],\label{eq:HoyosSonResult}
\eeq
where $\rho_0$ is the particle density, $\epsilon(B)$ is the energy density as a function of the external magnetic field $B$, $l_B=\sqrt{\hbar/|eB|}$ is the magnetic length of the system and $\omega_c=|eB|/m_e$ is the cyclotron frequency. According to Ref.~\cite{Biswas:2013wm}, for the integer QHE, the contribution proportional to $\epsilon''(B)$ may be interpreted as arising from the electric field-induced displacements of cyclotron orbit centres, while the contribution proportional to $\eta_H$ may be interpreted as arising from the electric field-induced shearing of the cyclotron orbits. A more general version of this quantity is derived in Ref.~\cite{Haldane:2015tx}, which separates the coefficient into diagonal and Landau-level-mixing contributions, allowing for broken rotational symmetry.

To simplify notation, we work in units where $\hbar=e=m_e=1$ and initially consider only isolated, filled Landau levels. In this scenario, we set $\nu=1$, and note that the Hall viscosity may be expressed in terms of the shift $\mcs$ as $\eta_H=\frac{1}{4}\hbar\rho_0\mathcal{S}$, where for the $n$th Landau level, $\mathcal{S}=2n+1$. The energy density of the $n$th Landau level is given by $\epsilon_n(B)=\hbar\omega_c(n+\frac{1}{2})/(2\pi l_B^2)$, since $2\pi l_B^2$ is the area of a cyclotron orbit. With these replacements, the Hall conductivity for the (isolated) $n$th Landau level takes the form
\beq
\sigma_{xyn}(q)&=&\frac{1}{2\pi}\left[1-\frac{3}{2}\left(n+\frac{1}{2}\right)\frac{q^2}{B}+O\left(\frac{q^4}{B^2}\right)\right].\label{eq:hall_cond_HS}
\eeq
In an experiment, one usually fills the lowest $K$ Landau levels. The Hall conductivity in this case can be found by summing the response from each filled band (i.e. $n=0,1,2,\ldots,K-1$) to obtain
\beq
\sigma_{xy}^K(q)&=&\frac{1}{2\pi}\left[K-\frac{3K^2}{4}\frac{q^2}{B}+O\left(\frac{q^4}{B^2}\right)\right].\label{eq:hall_cond_KLL}
\eeq

\subsection{Quantum Mechanical Derivation of Current Response\label{sec:QM_current_response}}
\begin{figure}[t]
\includegraphics[scale=0.5]{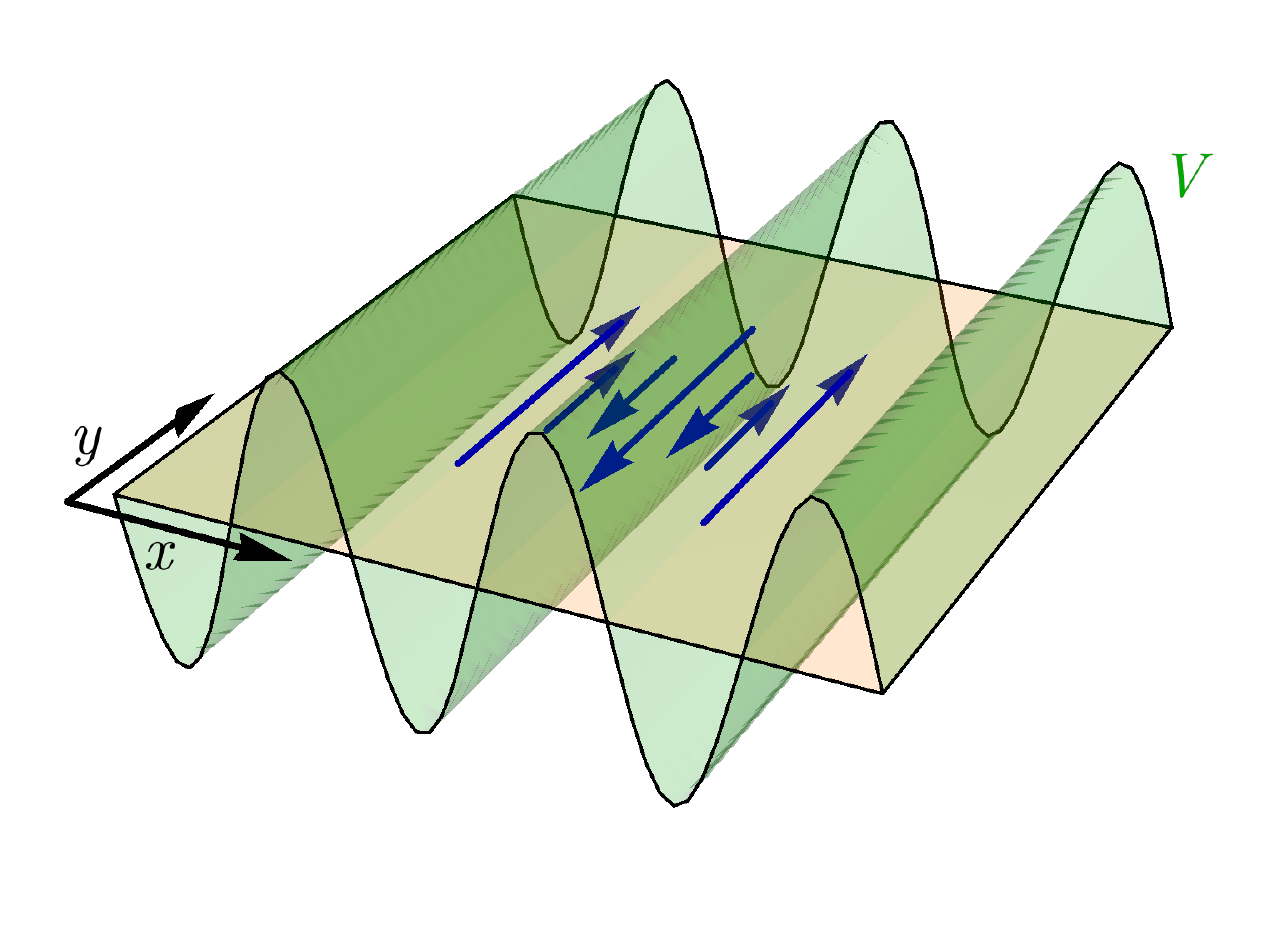}
\caption{Electrostatic potential in the $x$-direction $V(x)$ and the corresponding transverse current response $j_y$ (blue arrows). \label{fig:currdens_setup}}
\end{figure}
Our approach to the study of this quantity follows the setup of Ref.~\cite{Biswas:2013wm}, although the method we use will be different. We begin with the Landau Hamiltonian in the Landau gauge $\bA_L=B(0,x,0)$,
\beq
\h_L&=&\frac{1}{2}\left[\hat{p}_x^2+\left(\hat{p}_y-B\hat{x}\right)^2\right],
\eeq
where, as before, universal constants have been set to one. This Hamiltonian describes the motion of a charged particle in an external magnetic field of strength $B$, and may be solved in the standard way by defining the ladder operators
\beq
a&=&\frac{1}{\sqrt{2} l_B}\left(\hat{x}-k_yl_B^2\right)+i\frac{l_B}{\sqrt{2}}\hat{p}_x,\nonumber\\
a^\dagger&=&\frac{1}{\sqrt{2}l_B}\left(\hat{x}-k_yl_B^2\right)-i\frac{l_B}{\sqrt{2}}\hat{p}_x,\label{eq:hof_ladders}
\eeq
where we have inserted the magnetic length $l_B^2=1/B$. In terms of $a$ and $a^\dagger$, the Hamiltonian takes the form
\beq
\h_L&=&B\left(a^\dagger a^\ph+\frac{1}{2}\right),\label{eq:Landau_ham}
\eeq
with energy levels $E_n=\omega_c\left(n+\frac{1}{2}\right)$ (where $\omega_c=B$ in our units) and wavefunctions $\ket{n}=\left(a^\dagger\right)^n\ket{0}/\sqrt{n!}$, with $\ket{0}$ satisfying $a\ket{0}=0$. In real space, the wavefunctions are given by
\beq
\psi_{n,k_y}(\br)&=&\frac{e^{ik_yy}}{\sqrt{L_y}}\phi_n\left(x-k_y/B\right),\label{eq:LL_wavefunctions}
\eeq
with $\phi_n(x)$ a product of Gaussian and Hermite polynomial functions of $x$ \cite{landau1977quantum} and where $L_y$ is the system length in the $y$-direction.

We now apply a small inhomogeneous electrostatic potential to the system that acts only in the $x$ direction, $V(\hat{x})$ (see Fig.~\ref{fig:currdens_setup}). Since we are interested in the linear response of the system to this external potential, we assume that the magnitude of $V(\hat{x})$ is weak. In addition, we assume that $V(\hat{x})$ varies with some length scale $\lambda$ that is much larger than the characteristic length scale of the wavefunctions $l_B$. With these assumptions, we can treat $V(\hat{x})$ as a small perturbation to the Landau Hamiltonian, and expand it as a power series about some position $x_0$, where higher derivative terms become successively weaker. Specifically, we write
\beq
V(\hat{x})&=&\sum_{p=0}c_p\left(\hat{x}-x_0\right)^p,\label{eq:v_expand}
\eeq
with
\beq
c_p&=&\frac{1}{p!}\left.\frac{\partial^p V}{\partial x^p}\right|_{x=x_0}\sim\frac{1}{\lambda^p}.\label{eq:cp_coefficient}
\eeq

We will ultimately be interested in the response as a function of wavevector $q$, for which it is useful to use the explicit oscillating electric field in the $x$-direction
\beq
E(\hat{x})&=&E_qe^{iq\left(\hat{x}-x_0\right)}\left[=-\partial_xV(\hat{x})\right],
\eeq
where $E_q$ may be a complex number and we take the real part at the end of the calculation. For this choice of potential, the coefficients $c_m$ take the simple form
\beq
c_p&=&\frac{(iq)^{p-1}}{p!}E_q
\eeq
and we may identify $q=2\pi/\lambda$.

There are two meaningful current responses that we may calculate in this set up. The first is the (transverse) current per orbital, a single-particle quantity that describes the behaviour of an occupied single-particle orbital in the presence of the external field. The second is the (transverse) current density, a many-body property of the filled Landau level that includes contributions from all occupied cyclotron orbits at a given point in space. This latter quantity is directly connected to the Hall conductivity, $\sigma_{xy}$, and is related to the physical current that would be measured in an experiment. We calculate each of these quantities in turn below, treating the external potential as a weak perturbation. We note that this quantum mechanical calculation gives equivalent results to those that would be obtained using linear response theory and the Kubo formula, although the method is considerably simpler. The equivalence between these two approaches is shown explicitly in Appendix~\ref{app:j_linear_response} for the current density calculation.
\subsubsection{Current per Orbital}
We first calculate the perturbed orbital states that result from the action of the weak external potential, which enters the Hamiltonian through the term $\Delta H=V(\hat{x})$ (with $e=1$). From elementary perturbation theory, the first-order perturbed states are
\begin{equation}
\ket{\tilde{n},k_y}=\ket{n,k_y}+\sum_{m\neq n}\frac{\bra{m,k_y}V(\hat{x})\ket{n,k_y}}{E_n-E_m}\ket{m,k_y},
\end{equation}
where, since $V(\hat{x})$ acts only in the $x$-direction, $k_y$ is conserved. Expanding $V(\hat{x})$ about the centre of the orbital, $x_0=k_y/B$, we see that Eq.~\eqref{eq:v_expand} may be rewritten in terms of Landau level operators as
\beq
V(\hat{x})&=&\sum_pc_p\left(2B\right)^{-p/2}\left(a+a^\dagger\right)^p,
\eeq
allowing the perturbed wavefunctions to be calculated straightforwardly.

The operator that gives the current per orbital in the $y$-direction is obtained canonically through $\hat{I}_y=e\hat{v}_y=\partial \h_L/\partial k_y$, giving
\beq
\hat{I}_y=k_y-B\hat{x}=-\sqrt{\frac{B}{2}}\left(a+a^\dagger\right).
\eeq
The leading current response to the perturbation is equal to the expectation value of this operator evaluated in the perturbed orbital of interest, $\bra{\tilde{n},k_y}\hat{I}_y\ket{\tilde{n},k_y}$. The zeroth order term vanishes trivially, and the leading contribution is given by
\begin{equation}
\left\langle\hat{I}_{yn}\right\rangle=\sum_{m\neq n}\bra{n,k_y}\hat{I}_y\ket{m,k_y}\frac{\bra{m,k_y}V(\hat{x})\ket{n,k_y}}{E_n-E_m}+\hc\label{eq:curr_per_orb_1}
\end{equation}
After some algebra, we obtain
\beq
\left\langle\hat{I}_{yn}\right\rangle&=&-\frac{c_1}{B}-\frac{3c_3}{B^2}\left(n+\frac{1}{2}\right)-\frac{15c_5}{2B^3}\left(n^2+n+\frac{1}{2}\right)+\ldots\nonumber\\
&=&l_B^2\left[1+\frac{1}{2}\left(n+\frac{1}{2}\right)l_B^2\partial_x^2\right.\label{eq:current_orb_LL}\\
&&\left.\left.+\frac{1}{16}\left(n^2+n+\frac{1}{2}\right)l_B^2\partial_x^4+\ldots\right]E(x)\right|_{x=k_y/B},\nonumber
\eeq
where in the second line we have substituted for $c_p$ and used $E(x)=-\partial_xV(x)$ to write the expression in terms of the electric field. This is the current that would arise instantaneously if a single electron, in orbital $\ket{n,k_y}$, were suddenly exposed to a weak electric field described by $E(x)$. The magnitude and sign of this current depend on the spatial location of the orbital relative to the external field through $E(k_y/B)$, and for a harmonic potential would oscillate in space as in Fig.~\ref{fig:currdens_setup}. The expression above agrees with the current per orbital result given in Ref.~\cite{Biswas:2013wm}.

Finally, we note that the current per orbital is a band geometric quantity---i.e., it depends only on the wavefunctions of the system and not on the energies. To see this, we write $\hat{I}_y=\hat{v}_y=-i\left[\hat{y},\h_L\right]$ in Eq.~\eqref{eq:curr_per_orb_1} (implicitly assuming the system is infinite in order for $\hat{y}$ to be well defined), and note that the action of $\h_L$ cancels the difference of energies in the denominator to give
\begin{align}
\left\langle\hat{I}_{yn}\right\rangle&=\sum_{m\neq n}i\bra{n,k_y}\hat{y}\ket{m,k_y}\bra{m,k_y}V(\hat{x})\ket{n,k_y}+\hc\nonumber\\
&=-i\bra{n,k_y}\hat{y}\ket{n,k_y}\bra{n,k_y}V(\hat{x})\ket{n,k_y}+\hc,\label{eq:curr_per_orbital_BG}
\end{align}
where in the second line we have used $\sum_{m\neq n}\ket{m}\bra{m}=\id-\ket{n}\bra{n}$. Since $V(\hat{x})$ may be written as a series in $\hat{x}$, the expression above only depends on the algebra of the projected position operators, and is thus a geometric property of the bands. In particular, the first nonzero term is related to the Berry curvature of the band. 

Band geometric quantities are important in determining the stability of many-body FCI states built from the band eigenstates \cite{Roy:2014dl,Jackson:2015fv,Bauer:2016ju}. The equivalence in Eq.~\eqref{eq:curr_per_orbital_BG} shows that the current per orbital may provide a route to measuring the band geometry experimentally.
\subsubsection{Current Density}
We now consider the current density of a filled, isolated Landau level. This differs from the calculation above in two main ways. First, we require the current at a particular point in space rather than the total current of a cyclotron orbit. For a single particle, this corresponds to the current density operator,
\beq
\hat{j}_y(\br_0)&=&\frac{1}{2}\left[\hat{I}_y\delta(\hat{\br}-\br_0)+\delta\left(\hat{\br}-\br_0\right)\hat{I}_y\right].
\eeq
Secondly, we must sum over contributions to the current density from all states in the Landau level. In this case, we will sum over different values of $k_y$, which we assume take values $2\pi m/L_y$ with $m=0,1,\ldots,L_y-1$.

Following a similar line of argument to before, we perturb the Landau level states with the external potential, and calculate expectation value of the current density in these new states, arriving at the expression
\beq
J_{yn}(\br_0)&=&\sum_{k_y}\left[\sum_{m\neq n}\bra{n,k_y}\hat{j}_y(\br_0)\ket{m,k_y}\times\right.\nonumber\\
&&\left.\frac{\bra{m,k_y}V(\hat{x})\ket{n,k_y}}{E_n-E_m}+\hc\right],\label{eq:curr_dens_1}
\eeq
where the zeroth order term can again be shown to vanish. We emphasise that this formulation is equivalent to calculating the current density using many-body linear response theory, a correspondence that is shown explicitly in Appendix~\ref{app:j_linear_response}.

A further complication arises in how we expand the external potential. In order to consistently sum current contributions from each orbital, we must expand $V(\hat{x})$ about the same point in space in each case. We take this point to be $x_0$, the $x$-coordinate of the position at which we are measuring the current, and rewrite Eq.~\eqref{eq:v_expand} as
\beq
V(\hat{x})&=&\sum_{p}c_p\left[\left(\hat{x}-\frac{k_y}{B}\right)-\left(x_0-\frac{k_y}{B}\right)\right]^p,\nonumber\\
&\equiv&\sum_{r,s=0}d_{rs}\left(\hat{x}-\frac{k_y}{B}\right)^r\left(x_0-\frac{k_y}{B}\right)^s,\label{eq:v_expand2}
\eeq
where 
\beq
d_{rs}&=&(-1)^s \binom{r+s}{r} c_{r+s}.
\eeq
With this substitution, we have separated the operator-valued terms proportional to $\left(\hat{x}-\frac{k_y}{B}\right)^r=\left(\frac{a+a^\dagger}{\sqrt{2B}}\right)^r$, which act to perturb the Landau level states, from the scalar terms proportional to $\left(x_0-\frac{k_y}{B}\right)^s$, which will ultimately be integrated over. Substituting these and the explicit wavefunctions from Eq.~\eqref{eq:LL_wavefunctions} into Eq.~\eqref{eq:curr_dens_1}, we obtain
\begin{widetext}
\small
\begin{equation}
J_{yn}(\br_0)=\int \frac{\dd k_y}{2\pi}\sum_{r,s=0}\sum_{m\neq n}d_{rs}\left[\phi_n^*\left(x_0-\frac{k_y}{B}\right)\right]\left[{I}_y\left(x_0-\frac{k_y}{B}\right)\right]\left[\phi_m\left(x_0-\frac{k_y}{B}\right)\right]\left[\left(x_0-\frac{k_y}{B}\right)^s\right]\left[\frac{\bra{m,k_y}\left[\frac{a+a^\dagger}{\sqrt{2B}}\right]^r\ket{n,k_y}}{E_n-E_m}\right]+\hc,
\end{equation}
\normalsize
where we have also taken the continuum limit in replacing $\sum_{k_y}\to\int\dd k_y\frac{L_y}{2\pi}$. The final factor is independent of $k_y$, and so we can drop this label. For the remaining factors, we define the new variable $u=x_0-k_y/B$ and write the integral over $k_y$ as an inner product, arriving at
\begin{equation}
J_{yn}(\br_0)=\frac{B}{2\pi}\sum_{r,s=0}\sum_{m\neq n}d_{rs}\bra{n}\hat{I}_y\hat{x}^s\ket{m}\frac{\bra{m}\hat{x}^r\ket{n}}{E_n-E_m}+\hc
\end{equation}
\end{widetext}
In this way, in the thermodynamic limit, the current density reduces to a sum over products of matrix elements, which can be calculated out straightforwardly using ladder operators. After some algebra, we obtain
\beq
J_{yn}(\br_0)&=&-\frac{B}{2\pi}\left[\frac{c_1}{B}+\frac{9c_3}{B^2}\left(n+\frac{1}{2}\right)\right.\\
&&\left.+\frac{5c_5}{2B^3}\left(11+30n+30n^2\right)+\ldots\right],\nonumber
\eeq
or using Eq.~\eqref{eq:cp_coefficient},
\beq
J_{yn}(\br_0)&=&\frac{1}{2\pi}\left[1+\frac{3}{2}\left(n+\frac{1}{2}\right)l_B^2\partial_x^2\right.\label{eq:curr_dens_Ex}\\
&&\left.\left.+\frac{1}{48}\left(11+30n+30n^2\right)l_B^4\partial_x^4+\ldots\right]E(x)\right|_{x=x_0}.\nonumber
\eeq
Under the substitution $E(x)=E_qe^{iq\left(\hat{x}-x_0\right)}$, and using Eq.~\eqref{eq:j_sigma_e}, we identify the finite-wavevector Hall conductivity for the $n$th Landau level as
\beq
\sigma_{xyn}(q)&=&\frac{1}{2\pi}\left[1-\frac{3}{2}\left(n+\frac{1}{2}\right)\frac{q^2}{B}\right.\\
&&\left.+\frac{1}{48}\left(11+30n+30n^2\right)\frac{q^4}{B^2}+\ldots\right],\nonumber
\eeq
which agrees at $O(q^2)$ with the result from the literature in Eq.~\eqref{eq:hall_cond_HS}. We emphasise that this method can be extended (and automated) straightforwardly to obtain terms up to as high an order in $q$ as desired. The general coefficient will be a polynomial in the Landau level index $n$.

The current density is directly related to the physical current that would be measured in a transport experiment, with the precise relation depending on the nature of the probe. If the probe is small compared to the length scale of the potential $\lambda$, then the current measured at position $\br_0$ would be proportional to the integral of $J_{yn}(\br)$ over the probe area around $\br_0$. For a harmonic potential, the current density would oscillate in space according to the value of $E(x_0)$ in Eq.~\eqref{eq:curr_dens_Ex}, as shown heuristically in Fig.~\ref{fig:currdens_setup}. If the probe size is similar to the length scale of the potential, then the oscillating contributions to the current density may cancel out.

Unlike the current per orbital, the current density is not a band geometric quantity: the presence of the delta function in the current density operator prevents the energy denominators from being cancelled. However, the current density is much easier to measure in an experiment, as it requires filling all the states in a Landau level rather the selectively occupying certain orbitals.
\section{Perturbative Approach to Lattice Quantum Hall Systems\label{sec:perturbationtheory}}
\subsection{Overview}
We now turn our attention to quantum Hall physics on a lattice. Specifically, we consider topologically trivial tight-binding bands that become topological only when an external magnetic field is applied (although our method should also have applications to the study of Chern bands). An archetypal example of such a system is the Hofstadter model \cite{Harper:1955uu,Hofstadter:1976wt}, which describes charged particles hopping on a square lattice in the presence of a magnetic field. In the weak field limit, the eigenstates of the Hofstadter model connect smoothly to the Landau levels of the continuum \cite{Harper:2014fq}, while at stronger field strengths, the energy spectrum takes on the famous fractal butterfly structure. In this section, we argue that a similar, smooth connection to Landau level physics arises in generic lattice models.

In a tight-binding model, the effect of a magnetic field is usually included through the Peierls substitution \cite{Peierls:1933tv}. This is the addition of a complex phase to the hopping parameters that depends on the line integral of the vector potential along the shortest path between the two sites involved in the hop,
\beq
tc^\dagger_{\br'} c^\ph_{\br}&\longrightarrow&te^{-i\int_{\br}^{\br'}\bA\cdot\dd\mathbf{l}}c^\dagger_{\br'} c^\ph_{\br}.
\eeq
Although an approximation, the Peierls substitution describes the physics of the tight-binding model accurately in the limit where the vector potential varies slowly on the order of the lattice spacing \cite{Luttinger:1951bv,Kohn:1959zza,Blount:1962jn}.

The Hamiltonian for the Hofstadter model derives from a simple square lattice model in this way. Explicitly, beginning from the square-lattice hopping Hamiltonian
\begin{align}
\h_{\rm sq}&=-t\sum_\br\left[c^\dagger_{\br+\mathtt{a}\bx}c^\ph_{\br}+c^\dagger_{\br-\ta\bx}c^\ph_\br+c^\dagger_{\br+\ta\by}c^\ph_{\br}+c^\dagger_{\br-\ta\by}c^\ph_\br\right],
\end{align}
we implement the Peierls substitution in the Landau gauge ($\bA=B(0,x,0)$) to arrive at
\begin{align}
\h_{\rm Hof}&=-t\sum_\br\left[c^\dagger_{\br+\ta\bx}c^\ph_{\br}+c^\dagger_{\br-\ta\bx}c^\ph_\br\right.\label{eq:hofham}\\
&\left.+e^{-iBx\ta^2}c^\dagger_{\br+\ta\by}c^\ph_{\br}+e^{iBx\ta^2}c^\dagger_{\br-\ta\by}c^\ph_\br\right].\nonumber
\end{align}
In the above expressions, $t$ is the nearest-neighbour hopping parameter, $\br$ takes values on a square lattice with lattice spacing $\mathtt{a}$, $c^\dagger_\br$ is a fermionic creation operator at site $\br$ and 
\beq
B\ta^2\equiv\frac{\phi}{\phi_0}\equiv2\pi\phi
\eeq
is the magnetic flux per plaquette (the flux quantum $\phi_0=e^2/h=1/2\pi$ in our units).

\begin{figure}[t]
\includegraphics[scale=0.5]{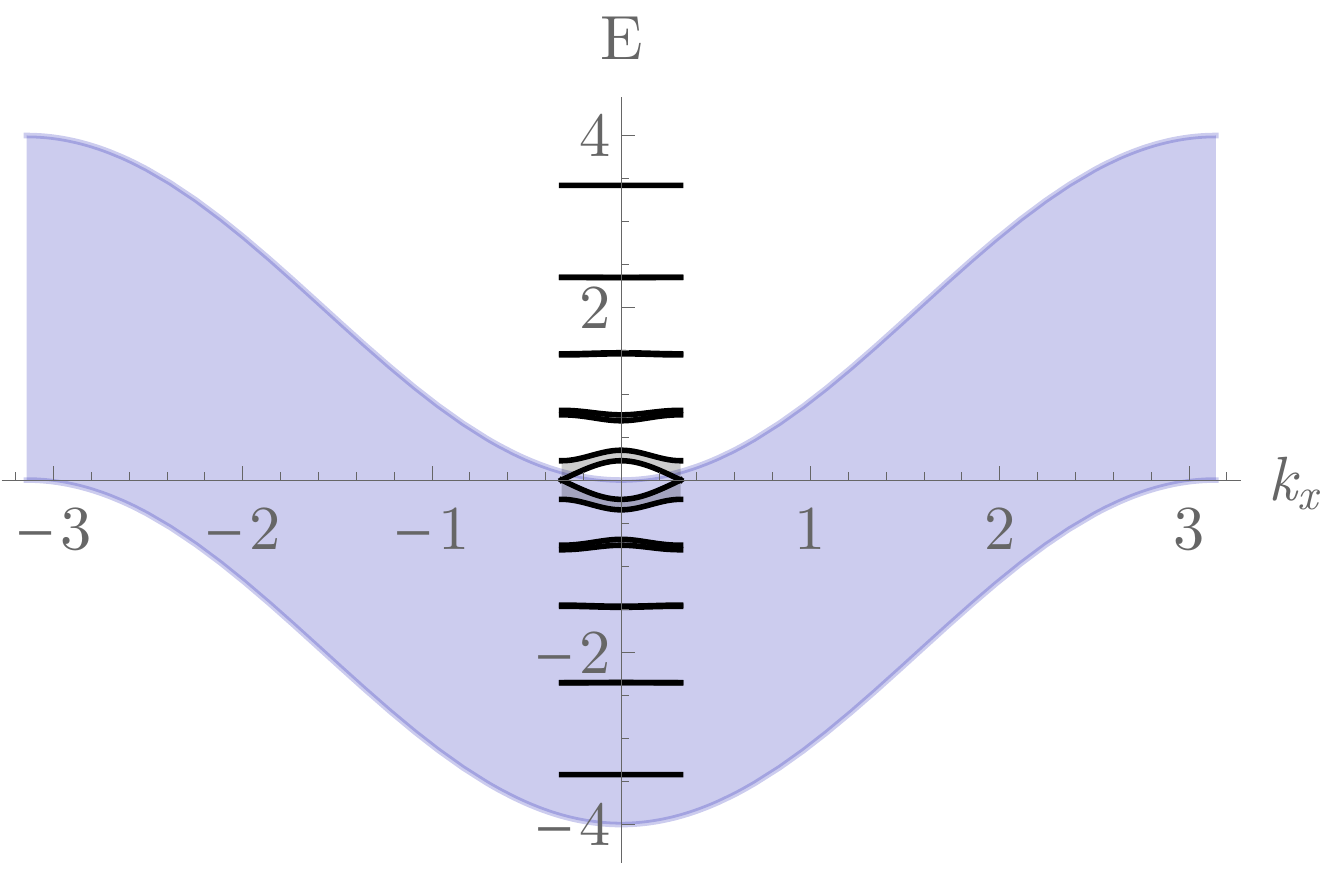}
\caption{Band structure of the zero-field square lattice model (large blue band) and the Hofstadter model at $\phi=1/10$ (ten black bands). The Hofstadter bands may be thought of as arising from a folding and splitting of the zero-field band structure. The low- and high-energy Hofstadter bands resemble flat Landau levels. \label{fig:spectrum110}}
\end{figure}

The presence of a magnetic field extends the size of the repeating unit cell of the system. For $\phi=p/q$, with $p$ and $q$ coprime, the magnetic unit cell is $q\times1$ plaquettes in size and contains a total of $p$ flux quanta. Correspondingly, the Brillouin zone extent is reduced by a factor of $q$ in the $k_x$-direction, and the resulting $q$ bands can be thought of as arising through the folding and mixing of the original band structure within the reduced magnetic Brillouin zone, as illustrated in Fig.~\ref{fig:spectrum110}. If $\phi$ is tuned by sweeping the magnetic field, the number of bands in the system changes rapidly as a function of $q$, repeating outside of the interval $\phi\in[0,1)$ and forming a Cantor set for irrational values of $\phi$ \cite{Wilkinson:1987tt}.

In this paper we will restrict the discussion to rational values of $\phi$ (since any real $\phi$ can be approximated to arbitrary accuracy by a rational number), and we will in addition assume that $\phi$ is small, writing $\phi=1/N$ with $N$ a large integer. In this limit, low- and high-energy bands of the Hofstadter model resemble continuum Landau levels (see, for example, the two lowest and two highest Hofstadter bands in Fig.~\ref{fig:spectrum110}). This connection was pursued in detail in Refs.~\cite{Harper:2014fq,Bauer:2016ju}, where deviations from Landau level physics were found to fall into two categories: exponentially small corrections, which vanish in the weak-field limit (for fixed lattice spacing) as $e^{-\sigma N}$ for some approximately constant $\sigma$, and perturbative corrections, which vanish in the weak-field limit as $(1/N^k)$ for positive integers $k$. 

These two types of contribution can be seen to arise as follows. Consider the action of the Hofstadter Hamiltonian, Eq.~\eqref{eq:hofham}, on a single tight-binding orbital, which we write as $\psi(x,y)$. Since $k_y$ is a good quantum number in the Landau gauge, wavefunction solutions factorise into a Bloch wave in the $y$-direction, and an $x$-dependent part that satisfies
\bequ
E\psi(x)=-t\left[e^{-\ta\partial_x}+e^{\ta\partial_x}+2\cos\left(\frac{2\pi x\ta}{N}-\ta k_y\right)\right]\psi(x),\label{eq:Harper}
\eequ
an equation known as Harper's equation \cite{Harper:1955uu}. Although this equation is formally discrete, we can approximate it as a continuum equation, and expand each term on the right-hand side order by order in $\ta$. At first order, we find
\beq
\left(\frac{E+4t}{2}\right)\psi(x)&=&-\frac{t\ta^2}{2}\left[\partial_x^2-\frac{1}{l_B^4}\left(x-k_yl_B^2\right)^2\right]\psi(x)\nonumber\\
&\equiv& t\left(\frac{\ta}{l_B}\right)^2\left(a^\dagger a+\frac{1}{2}\right)\psi(x),
\eeq
where in the second line we have identified $1/l_B^2=2\pi\phi\equiv B\ta^2$ and used the ladder operators as defined in Eq.~\eqref{eq:hof_ladders}. The Hamiltonian above resembles the Landau Hamiltonian from Eq.~\eqref{eq:Landau_ham}, and this mapping to the continuum can be made exact if we identify 
\beq
t&=&\frac{\omega_c}{2}\left(\frac{l_B}{\ta}\right)^2.
\eeq

In this way, wavefunctions of the Hofstadter model may be approximated at lowest order by Landau level wavefunctions. Corrections to these may be obtained by expanding Eq.~\eqref{eq:Harper} to higher order, and treating the successive terms as perturbations \cite{Harper:2014fq}. Since higher terms in the series expansion arise at higher powers of $(\ta/l_B)$, these perturbations are small in the limit that the magnetic length is much larger than the lattice spacing. For a fixed lattice, this regime may be obtained by making the applied magnetic field very small.

These perturbative corrections account for the algebraic deviations from Landau level physics, vanishing as $B^k$ in the weak-field limit with fixed lattice spacing. Nonperturbative corrections arise due to tunnelling and discreteness effects, as may be seen heuristically from the structure of Eq.~\eqref{eq:Harper}: The cosine term on the right-hand side may be expanded about any of its minima, which are separated by a distance $x\ta =N$. The true wavefunction should include tunnelling between Landau level-like wavefunctions that reside in each of these minima. The tunnelling should be exponentially small in the barrier width, which in this case would lead to contributions that are of approximate size $e^{-\sigma N}$. A more careful consideration of tunnelling effects using the WKB approximation was carried out in Ref.~\cite{Harper:2014fq}, confirming these rough arguments. 

For a weak enough magnetic field, the exponentially small corrections will be negligible compared to the perturbative corrections, and so may be ignored. Specifically, this amounts to ignoring the dispersion of the energy bands and the exponentially small corrections to the wavefunctions (which in turn would affect band geometric properties such as the Berry curvature and quantum metric). In the next subsection, we will calculate the dominant perturbative corrections of this form for a generic lattice model, and identify the effect these have on the energy levels, wavefunctions and transport properties of the system.
\subsection{Bravais Lattices}
In the absence of a magnetic field, a generic term in a tight-binding model on a Bravais lattice may be written
\beq
\h_{mn}^0&=&-t_{mn}\sum_\br c^\dagger_{\br+m\bx+n\by}c^\ph_\br+\hc,
\eeq
where, to simplify notation, we set the lattice spacing $\ta=1$ from now on. Taking the Fourier transform, this Hamiltonian may be diagonalised in momentum space as
\beq
\h_{mn}^0(\bk)&=&-2t_{mn}\sum_\bk\cos\left(mk_x+nk_y\right)c^\dagger_\bk c^\ph_\bk,\label{eq:hmn_nop}
\eeq
where $k_x$ and $k_y$ take values between $-\pi$ and $\pi$. For simplicity, we have assumed that $t_{mn}$ is real, but the method may be extended straightforwardly to complex hopping parameters. 

We now introduce a magnetic field (in the Landau gauge) through the Peierls substitution, which changes the hopping term to
\bequ
\h_{mn}=-t_{mn}\sum_\br c^\dagger_{\br+m\bx+n\by}c^\ph_\br e^{-iB\left(xn+\frac{mn}{2}\right)}+\hc
\eequ
Motivated by the discussion in the previous subsection, we consider the action of this hopping term on a localised tight-binding orbital $\psi(x,y)$ and use the fact that $k_y$ is a good quantum number to obtain a discrete difference equation for just the $x$-dependent part, $\psi(x)$,
\beq
\h_{mn}\psi(x)&=&-t_{mn}\left[e^{ik_yn}e^{-iB\left(xn+\frac{mn}{2}\right)}e^{m\partial_x}\right.\\
&&\left.+e^{-ik_yn}e^{iB\left(xn-\frac{mn}{2}\right)}e^{-m\partial_x}\right]\psi(x).\nonumber
\eeq
Finally, we use the Baker-Campbell-Hausdorff identity to rewrite the above expression as
\beq
\h_{mn}\psi(x)&=&-2t_{mn}\cos\left[m\left(-i\partial_x\right)+n\left(k_y-Bx\right)\right]\psi(x)\nonumber\\
&\equiv&-2t_{mn}\cos\left[m\hat{k}_x+n\hat{k}_y\right]\psi(x).\label{eq:hmn_op}
\eeq
By comparing Eqs.~\eqref{eq:hmn_nop} and \eqref{eq:hmn_op}, we see that the real-space finite-field Hamiltonian may be obtained from the zero-field Hamiltonian through the substitution $\bk\to\hat{\bk}$, with $\hat{\bk}$ defined through Eq.~\eqref{eq:hmn_op}. In other words, the lattice Peierls substitution is equivalent to applying the minimal coupling substitution $\bp\to\bp-\bA$ at all orders in the Hamiltonian (this may be verified for other choices of gauge).

A complete tight-binding Hamiltonian will usually contain a sum of many hopping terms of the form $\h_{mn}$. From the discussion above, it follows that each term $\h_{mn}$ contributes a cosine operator as in Eq.~\eqref{eq:hmn_op}, which may be thought of as deriving from a zero-field term $\h_{mn}^0(\bk)$ through the substitution $\bk\to\hat{\bk}$. Notably, since each cosine term can be expanded as a power series, and terms from each power series recombined, the substitution $\bk\to\hat{\bk}$ may be applied directly to the zero-field band structure, which might only be known to low orders in $\bk$ from experiment or numerical results. This motivates a prescription for obtaining the finite-field Hamiltonian from a zero-field band structure:
\begin{itemize}
\item	Take $E(\bk)$ or $\h(\bk)$ for the zero-field tight-binding model on a Bravais lattice and substitute $k_x\to\hat{k}_x$ and $k_y\to\hat{k}_y$.
\item	Products should be replaced with their fully symmetrised operator equivalents, through
\beq
k_x^mk_y^n&\to&\frac{\{\hat{k}_x^m\hat{k}_y^n\}}{\binom{m+n}{m}},\label{eq:operator_ordering}
\eeq
where $\{\cdots\}$ indicates the fully symmetrised sum over distinct orderings.
\end{itemize}
This latter condition arises from considering the operator ordering in the expansion of each cosine term. For example, a zero-field band structure contribution proportional to $k_x^3k_y^1$ should be replaced with
\beq
k_x^3k_y&\to&\left(\hat{k}_x\hat{k}_x\hat{k}_x\hat{k}_y+\hat{k}_x\hat{k}_x\hat{k}_y\hat{k}_x\right.\nonumber\\
&&\left.+\hat{k}_x\hat{k}_y\hat{k}_x\hat{k}_x+\hat{k}_y\hat{k}_x\hat{k}_x\hat{k}_x\right)/4.
\eeq

Provided all terms of the series expansion are kept (or if each term of the Hamiltonian is written in cosine form), the substitution above is exact. To obtain a perturbative solution, we follow the approach outlined in the previous section and truncate the series order by order in $\bk$. As a specific example, we consider a tight-binding model on a Bravais lattice with $C_4$ symmetry (i.e. symmetry under $(x\to y, y\to-x)$). The most general zero-field band structure for such a model may be written
\begin{align}
E_{C_4}(\bk)&=C_{0,0}^0+C^2_{2,0}\left(k_x^2+k_y^2\right)+C^4_{4,0}\left(k_x^4+k_y^4\right)\label{eq:C4bandstructure}\\
&+C^4_{3,1}\left(k_x^3k_y-k_xk_y^3\right)+C^4_{2,2}k_x^2k_y^2+\ldots,\nonumber
\end{align}
where we have kept all allowable terms up to quartic order. At lowest order, we assume that $C^2_{2,0}$ is nonzero and apply the substitution $\bk\to\hat{\bk}$ to obtain
\beq
\h^{(1)}_{C_4}&=&C_{0,0}^0+C^2_{2,0}\left(-\partial_x^2+\left(k_y-Bx\right)^2\right).\label{eq:C4Ham1}
\eeq
By defining the usual ladder operators, which we may rewrite as
\beq
a&=&\frac{i}{\sqrt{2B}}\left(\hat{k}_x+i\hat{k}_y\right)\label{eq:k_ladders}\\
a^\dagger&=&-\frac{i}{\sqrt{2B}}\left(\hat{k}_x-i\hat{k}_y\right),\nonumber
\eeq
we obtain
\beq
\h^{(1)}_{C_4}&=&C_{0,0}^0+2BC^2_{2,0}\left(a^\dagger a+\frac{1}{2}\right).
\eeq
Thus, in a weak magnetic field, the tight-binding wavefunctions resemble Landau level wavefunctions with cyclotron frequency $\omega_c=2BC^2_{2,0}$ and overall energy offset $C^0_{0,0}$.

The leading lattice corrections arise due to the quartic terms
\beq
\h^{(2)}_{C_4}&=&C^4_{4,0}\left(\kx\kx\kx\kx+\ky\ky\ky\ky\right)\label{eq:C4Ham2}\\
&&+\frac{C^4_{2,2}}{6}\left(\left\{\kx\kx\ky\ky\right\}\right)\nonumber\\
&&+\frac{C^4_{3,1}}{4}\left(\left\{\kx\kx\kx\ky\right\}-\left\{\kx\ky\ky\ky\right\}\right),\nonumber
\eeq
also given in terms of ladder operators in Appendix~\ref{app:c4symmetry}. This can be used to calculate perturbative corrections to the energy levels and wavefunctions using elementary perturbation theory, results that are given in Appendix~\ref{app:c4symmetry}. Higher order corrections can be calculated in a similar manner.

In this way, given a zero-field band structure, it is possible to read off a set of perturbed Landau levels that resemble the low-lying single-particle states that would arise in the presence of a weak magnetic field. The energy levels of these states are shifted relative to the unperturbed values, and the wavefunctions pick up corrections that cause them to adopt the symmetry of the lattice. In turn, these wavefunctions may be used in the calculation of transport properties, for example, which we pursue in the next section. 

The approach outlined above applies directly to Bravais lattices with other symmetries. In general, the zero-field band structure should be expanded about the momentum at minimum energy $\bk_0$ (which may not be at $\bk_0=0$), and the substitution $(\bk-\bk_0)\to\hat{\bk}$ should be enacted. Since the expansion is about a band extremum, the leading $\hat{\bk}$-dependence will be quadratic\footnote{We leave a discussion of systems where the leading quadratic term vanishes to future work.} and can always be solved to give a Landau level solution. Higher order terms will in general be more complicated than in the $C_4$-symmetric case, but the perturbative method remains the same. However, if the lattice has additional structure (through a sublattice, orbital or spin degree of freedom, for example), the method must be altered slightly. We discuss these cases in the next subsection.

As suggested previously, this perturbative approach has a number of limitations. First, it neglects the exponentially small corrections that arise due to discreteness and tunnelling effects. In addition, the perturbed wavefunctions that are obtained using this method are not Bloch periodic. Instead, they are quasilocal states that would need to be superposed into a Bloch wavefunction by hand (see Ref.~\cite{Harper:2014fq} for an example of how to do this). Finally, in order to consistently truncate the perturbation series at a given order, the magnetic field must be weak enough that the neglected terms are negligible. Nevertheless, this method is expected to provide a good approximation across a wide regime of lattice parameters.
\subsection{Bravais Lattices with Substructure}
The majority of interesting tight-binding models have some sort of substructure even in the absence of a magnetic field---usually a combination of spin, orbital, and sublattice degrees of freedom. In such cases, a generic hopping term will take the form
\beq
\h_{mn\alpha\beta}^0&=&-t_{mn}^{\alpha\beta}\sum_\br c^\dagger_{\br+m\bx+n\by,\alpha}c^\ph_{\br,\beta}+\hc,\label{eq:orbital_hop}
\eeq
where $\alpha$ and $\beta$ now give the additional quantum numbers of the tight-binding orbital. A complete Hamiltonian will involve a sum of such terms and, by taking the Fourier transform, may be written in terms of the momentum space Hamiltonian matrix,
\beq
\h^0(\bk)&=&\sum_\bk\sum_{\alpha\beta}c^\dagger_{\bk,\alpha}\h_{\alpha\beta}^0(\bk)c^\ph_{\bk,\beta}.\label{eq:zf_matrix_ham}
\eeq
Note that there are a number of (gauge) choices for how to carry out this Fourier transform. In our case, the relevant matrix $\h_{\alpha\beta}^0(\bk)$ is the one that defines the periodic part of the Bloch wavefunction, $u_{\bk\gamma}(\br)$. This amounts to taking into account the spatial embedding of each orbital within the unit cell, through $u_{\bk\gamma}(\br)=e^{-i\bk\cdot\hat{\br}}\psi_{\bk\gamma}(\br)$, where $\psi_{\bk\gamma}(\br)$ is the Bloch periodic wavefunction and $\hat{\br}$ gives the spatial displacement of each orbital \cite{Bergholtz:2013ey}.

In the presence of a magnetic field, Eq.~\eqref{eq:orbital_hop} becomes
\bequ
\h_{mn\alpha\beta}=-t_{mn}^{\alpha\beta}\sum_\br e^{i\int_{\br+\delta\br_\beta}^{\br'+\delta\br_\alpha}}c^\dagger_{\br',\alpha}c^\ph_{\br,\beta}+\hc,
\eequ
where $\br'=\br+m\bx+n\by$ and $\delta\br_\alpha$ gives the spatial displacement of each orbital relative to the centre of the unit cell. Using the arguments of the previous section, we can consider the action of $\h_{mn\alpha\beta}$ on a localised tight-binding orbital $\psi_\beta(x,y)$ and use the fact that, in the Landau gauge, $k_y$ is well defined. As before, we find that the finite-field Hamiltonian may be derived from the zero-field Hamiltonian through the substitution $\bk\to\hat{\bk}$ (subject to the operator ordering given in Eq.~\eqref{eq:operator_ordering}). In other words, the new Hamiltonian is the matrix 
\beq
\h_{\alpha\beta}(\hat{\bk})&\equiv&H^0_{\alpha\beta}(\bk\to\hat{\bk}),
\eeq
considered to act on the real space wavefunctions $\psi_\beta(x,y)$.

Each matrix element $\h_{\alpha\beta}(\hat{\bk})$ can be expanded as a power series in $\bk$, and by solving the resulting matrix equation at each order, perturbative energy levels and wavefunctions may be obtained. Alternatively, the zero-field matrix elements $\h^0_{\alpha\beta}({\bk})$ may be expanded as a power series first, and then the (ordered) substitution $\bk\to\hat{\bk}$ carried out. The resulting wavefunctions will now be perturbed Landau levels with generally different amplitudes on each orbital.

In the Bravais lattice case, we were able to apply the substitution $\bk\to\hat{\bk}$ directly to the zero-field band structure, rather than to the Hamiltonian. This would not seem to apply in this case, as may be seen from considering the zero-field Hamiltonian in Eq.~\eqref{eq:zf_matrix_ham}. To obtain energy bands from this matrix, we need to solve a determinant equation and, in the process, find the root of a polynomial equation. Carrying out this procedure for the finite-field, operator-valued matrix changes the operator ordering in a way that depends on the original matrix elements, and which cannot be reverse engineered without knowing the full structure of the matrix. For example, expanding the square root
\beq
\sqrt{1+\hat{k}_x^2+\hat{k}_y^2}&=& 1+\frac{1}{2}\left(\hat{k}_x^2+\hat{k}_y^2\right)\\
&&-\frac{1}{8}\left(\hat{k}_x^4+\hat{k}_x^2\hat{k}_y^2+\hat{k}_y^2\hat{k}_x^2+\hat{k}_y^4\right)+\ldots,\nonumber
\eeq
generates the quartic terms $\hat{k}_x^2\hat{k}_y^2$ and $\hat{k}_y^2\hat{k}_x^2$, but none of the other four orderings. Given a zero-field band structure with a term $k_x^2k_y^2$, the usual symmetric replacement over all orderings does not apply. In this way, the substitution $\bk\to\hat{\bk}$ can only be applied consistently to the complete zero-field tight-binding matrix. Another way of stating this is that many different tight-binding models can give rise to the same energy bands, but the wavefunctions generally depend on the details of the model.

However, for certain symmetric models, this problem may be `accidentally' circumvented at first order, and the substitution $\bk\to\hat{\bk}$ may be successfully applied even to the zero-field band structure. For lattices with $C_4$ symmetry, the first problematic term in the band structure is $k_x^2k_y^2$. Although we do not know the correct operator substitution for the reasons mentioned above, we know that it must be symmetric under the simultaneous replacements $\hat{k}_x\to\hat{k}_y$ and $\hat{k}_y\to-\hat{k}_x$. We can therefore write
\beq
k_x^2k_y^2&\to&c_1\left(\hat{k}_x\hat{k}_y\hat{k}_y\hat{k}_x+\hat{k}_y\hat{k}_x\hat{k}_x\hat{k}_y\right)\\
&&+c_2\left(\hat{k}_x\hat{k}_y\hat{k}_x\hat{k}_y+\hat{k}_y\hat{k}_x\hat{k}_y\hat{k}_x\right)\nonumber\\
&&+\left(\frac{1}{2}-c_1-c_2\right)\left(\hat{k}_x\hat{k}_x\hat{k}_y\hat{k}_y+\hat{k}_y\hat{k}_y\hat{k}_x\hat{k}_x\right),\nonumber
\eeq
which, in terms of ladder operators, is
\beq
k_x^2k_y^2&\to&\frac{B^2}{4}\left[4a^\dagger a+2\ad{2}\an{2}\right.\\
&&\left.-\ad{4}-\an{4}+8c_1+4c_2-1\right].\nonumber
\eeq
Since the unknown coefficients only enter as constant offsets, the naive symmetric substitution $\bk\to\hat{\bk}$ will always have the correct operator dependence. In first order perturbation theory, this is enough to calculate the perturbed wavefunctions and the perturbed energy differences between states, all that we will require in the next section when we calculate the leading perturbations to the conductivity. For $C_4$ symmetric lattices, therefore, the perturbed solutions given for a Bravais lattice in Appendix~\ref{app:c4symmetry} apply also to models with a unit cell substructure (other than the overall energy offset). Similar simplifications may arise for other symmetries, but in general the perturbative method needs to be applied at the level of the Hamiltonian matrix. Conversely, measuring behaviour such as transport properties, in conjunction with a measured band structure, may allow the reverse engineering of an effective underlying tight-binding model for an experimental system.
\section{Electromagnetic Response of Lattice Quantum Hall Systems\label{sec:em_lattice}}
In the previous two sections, we showed how to calculate the finite-wavevector current response of a filled Landau level, and demonstrated that Landau-level like wavefunctions arise naturally in generic lattice models in the presence of a weak magnetic field. We now bring both of these strands together and calculate the leading corrections to the current response that arise due to the presence of a lattice. These could be important if, say, one wants to measure the Hall viscosity through a conductivity experiment.

The setup will be similar to that used in Sec.~\ref{sec:em_continuum}: we take a lattice quantum Hall system and apply a weak inhomogeneous electric field as shown in Fig.~\ref{fig:currdens_setup}. There are now three length scales in the system: the length scale of the potential $\lambda$, the magnetic length $\l_B=1/\sqrt{B}$, and the lattice spacing $\ta$. We are interested in the limit $\lambda\gg l_B\gg\ta$, where the discreteness of the lattice may be treated as a perturbation to the Landau level wavefunctions, and where the external potential may be truncated as in Eq.~\eqref{eq:v_expand} and used in linear response.

Applying the methods of Secs.~\ref{sec:em_continuum} and \ref{sec:perturbationtheory}, the single-particle wavefunctions may be expanded as a double perturbation series. The perturbation due to the lattice may be written
\beq
\ket{n',k_y}&=&U^\dagger\left(a,a^\dagger\right)\ket{n,k_y},
\eeq
where $\ket{n,k_y}$ is an unperturbed Landau level and $U^\dagger\left(a,a^\dagger\right)$ is a unitary operator that applies the perturbations described in Sec.~\ref{sec:perturbationtheory} (see Appendix~\ref{app:c4symmetry} for an explicit expression for this operator). These lattice wavefunctions are in turn perturbed by the external potential as described in Sec.~\ref{sec:em_continuum}, yielding
\beq
\ket{\tilde{n}',k_y}&=&\ket{n',k_y}+\sum_{m'\neq n'}\frac{\bra{m',k_y}V(\hat{x})\ket{n',k_y}}{E_n'-E_m'}\ket{m',k_y}\nonumber\\
&=&U^\dagger\ket{n,k_y}\\
&&+\sum_{m\neq n}\frac{\bra{m,k_y}UV(\hat{x})U^\dagger\ket{n,k_y}}{E_n'-E_m'}U^\dagger\ket{m,k_y}.\nonumber
\eeq
Since we are considering the linear response, we keep only the first order perturbation due to the potential; the lattice corrections may be applied up to as high an order as desired by truncating the resulting perturbation series appropriately.

There are again two meaningful transverse current responses that we can calculate: the current per single-particle orbital and the current density. However, we must now include corrections to the operator $\hat{I}_y$ due to the presence of the lattice. Using the definition $\hat{I}_y=\partial\h/\partial k_y$, we see that for a $C_4$ symmetric lattice, whose Hamiltonian takes the general form given in Eqs.~\eqref{eq:C4Ham1} and \eqref{eq:C4Ham2}, the current operator becomes
\beq
\hat{I}_y'&=&2C^2_{2,0}\ky+4C^4_{4,0}\ky\ky\ky+\frac{2C^4_{2,2}}{3}\left\{\kx\kx\ky\right\}\nonumber\\
&&+C^4_{3,1}\left(\kx\kx\kx-\left\{\kx\ky\ky\right\}\right).\label{eq:lat_current_momentum}
\eeq
This expression is given in terms of ladder operators in Appendix~\ref{app:c4symmetry}.
\subsection{Current Per Orbital\label{sec:lattice_curr_orb}}
The current per orbital is again given by the expectation value of this (now perturbed) current operator in a single-particle state, $\bra{\tilde{n}',k_y}\hat{I}_y'\ket{\tilde{n}',k_y}$, with leading non-zero terms
\begin{equation}
\left\langle\hat{I}_{yn'}'\right\rangle=\sum_{m'\neq n'}\bra{n',k_y}\hat{I}_y'\ket{m',k_y}\frac{\bra{m',k_y}V(\hat{x})\ket{n',k_y}}{E_n'-E_m'}+\hc\label{eq:curr_per_orb_lat}
\end{equation}
Lattice corrections enter this expression in three places: in the perturbed wavefunctions $\ket{n',k_y}$; in the perturbed energy denominator $E_n'-E_m'$; and through the perturbed current operator $\hat{I}_y'$. In each case, the corrections may be expressed as a perturbation series in powers of $B\equiv2\pi/N$, which is assumed to be small. A consistent expression for the current per orbital is obtained by collecting together terms from each source order by order.

For a $C_4$-symmetric lattice, the current per orbital up to first order in $B$ is 
\begin{widetext}
\beq
\left\langle\hat{I}_{yn'}^{C_4}\right\rangle&=&\left\langle\hat{I}_{yn}\right\rangle+Bl_B^2\left.\left[\frac{\left(C^4_{2,2}-2C^4_{4,0}\right)}{192C_{2,0}^{2}}\left(1+2n\right)\left(3+n+n^2\right)l_B^2\partial_x^4+\ldots\right]E(x)\right|_{x=k_y/B},\label{eq:curr_orb_c4}
\eeq
\end{widetext}
where $\left\langle\hat{I}_{yn}\right\rangle$ is the current per orbital for the unperturbed $n$th Landau level, given in Eq.~\eqref{eq:current_orb_LL}. Notably, the zeroth order (homogeneous) term is related to the quantised Hall conductance, and does not gain corrections at any order in perturbation theory. The second term, proportional to $\partial_x^2 E(x)$, is unchanged at first order, but it does pick up corrections at higher orders in perturbation theory. The corrections for the specific case of the Hofstadter model are given in Appendix~\ref{app:c4symmetry}.
\subsection{Current Density\label{sec:lattice_curr_dens}}
As in the continuum, the transverse current density is related to the transport response that would be measured in a real experiment. Following the approach of Sec.~\ref{sec:QM_current_response}, we calculate this quantity for a filled band by taking the expectation value of the current density operator and summing over all relevant single-particle states. This can be shown to be equivalent to a linear response calculation using the methods in Appendix~\ref{app:j_linear_response}.

The current density operator is now
\beq
\hat{j}_y'(\br_0)&=&\frac{1}{2}\left[\hat{I}'_y\delta(\hat{\br}-\br_0)+\delta\left(\hat{\br}-\br_0\right)\hat{I}_y'\right],
\eeq
which includes perturbative corrections due to the lattice as described above. We take the expectation value of this operator with respect to the single particle states $\ket{\tilde{n}',k_y}$, which have been perturbed by both the lattice and the external potential. Summing over all the states in the perturbed band, we find
\begin{equation}
J_{yn}'(\br_0)=\frac{B}{2\pi}\sum_{r,s=0}\sum_{m'\neq n'}d_{rs}\bra{n'}\hat{I}_y'\hat{x}^s\ket{m'}\frac{\bra{m'}\hat{x}^r\ket{n'}}{E_n'-E_m'}+\hc
\end{equation}
where, as in the continuum, the final expression is a sum over a product of matrix elements, which may be calculated straightforwardly using ladder operators. Lattice corrections enter through the perturbed wavefunctions, energies and current operator.

\begin{figure*}[t!]
\includegraphics[scale=0.45]{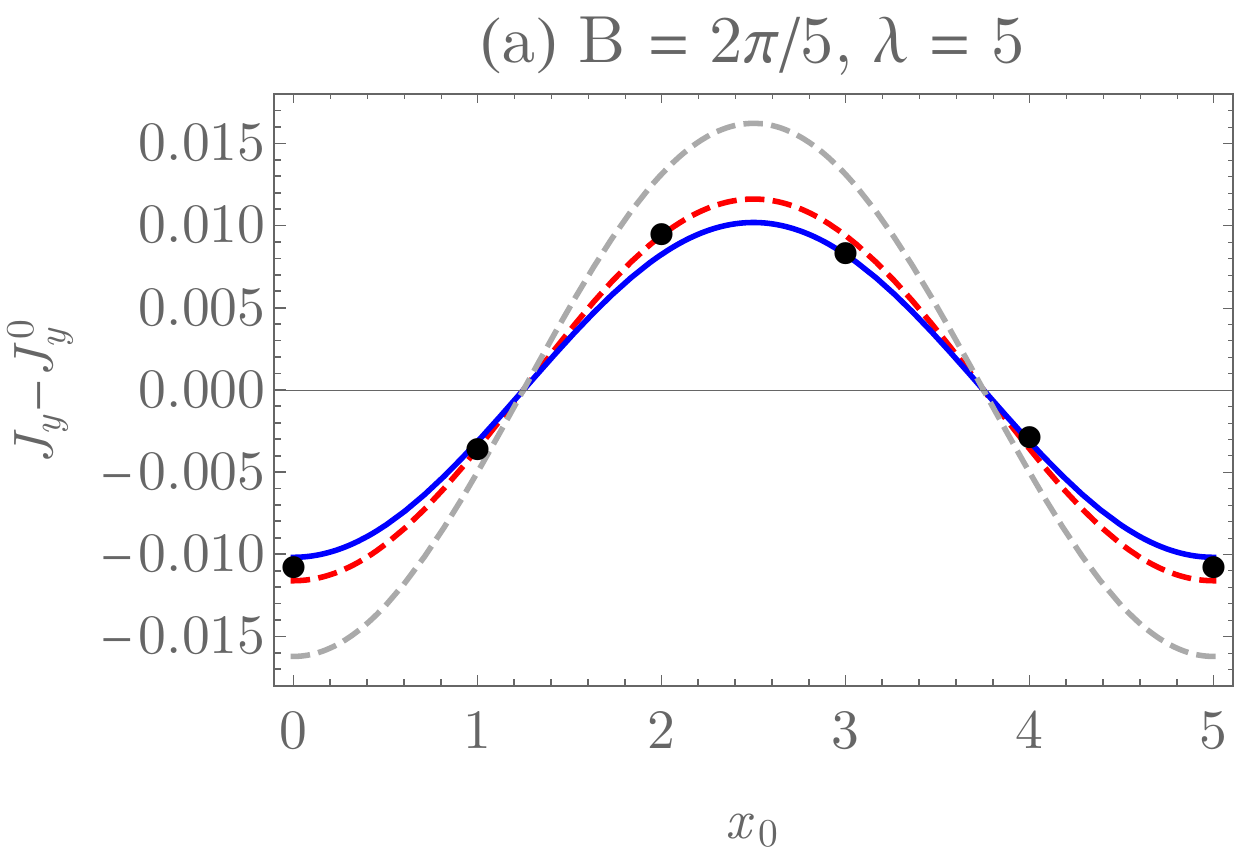}
\includegraphics[scale=0.45]{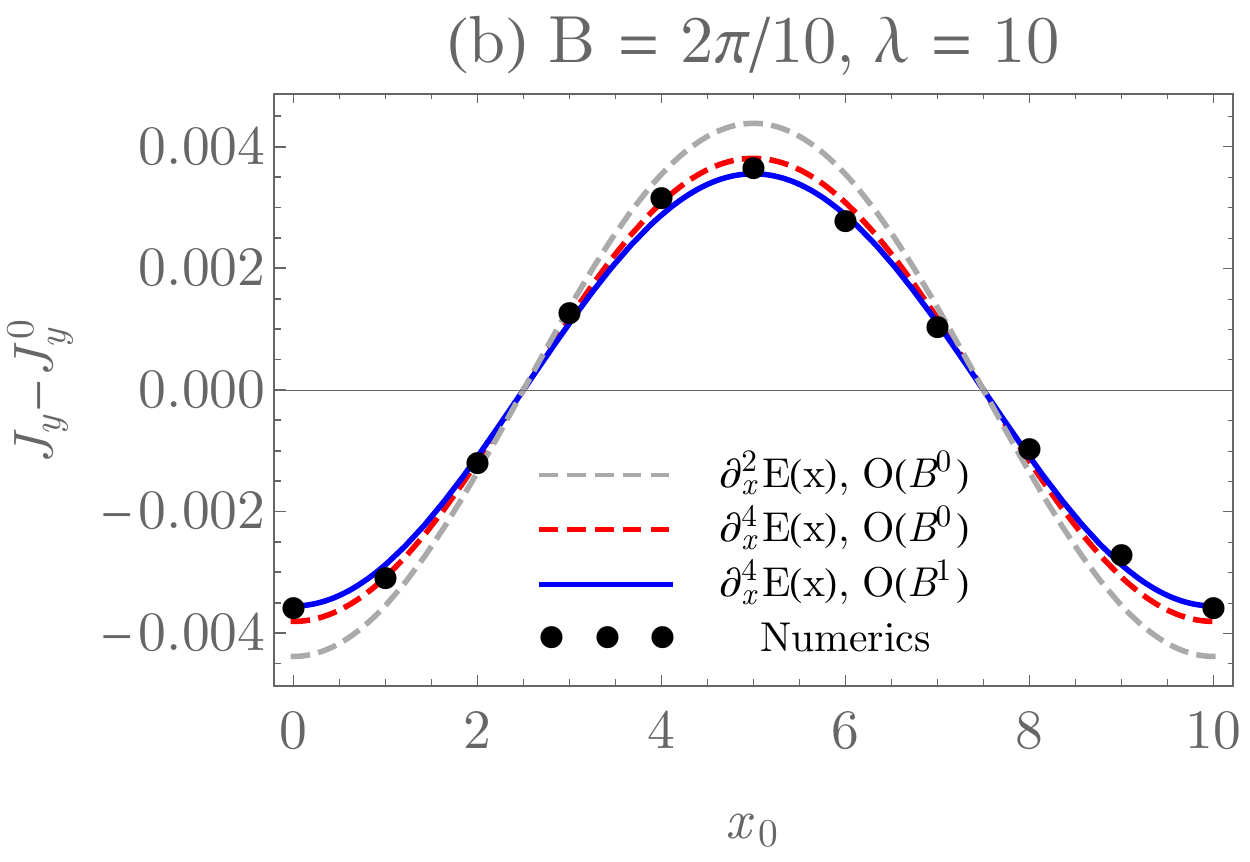}
\includegraphics[scale=0.45]{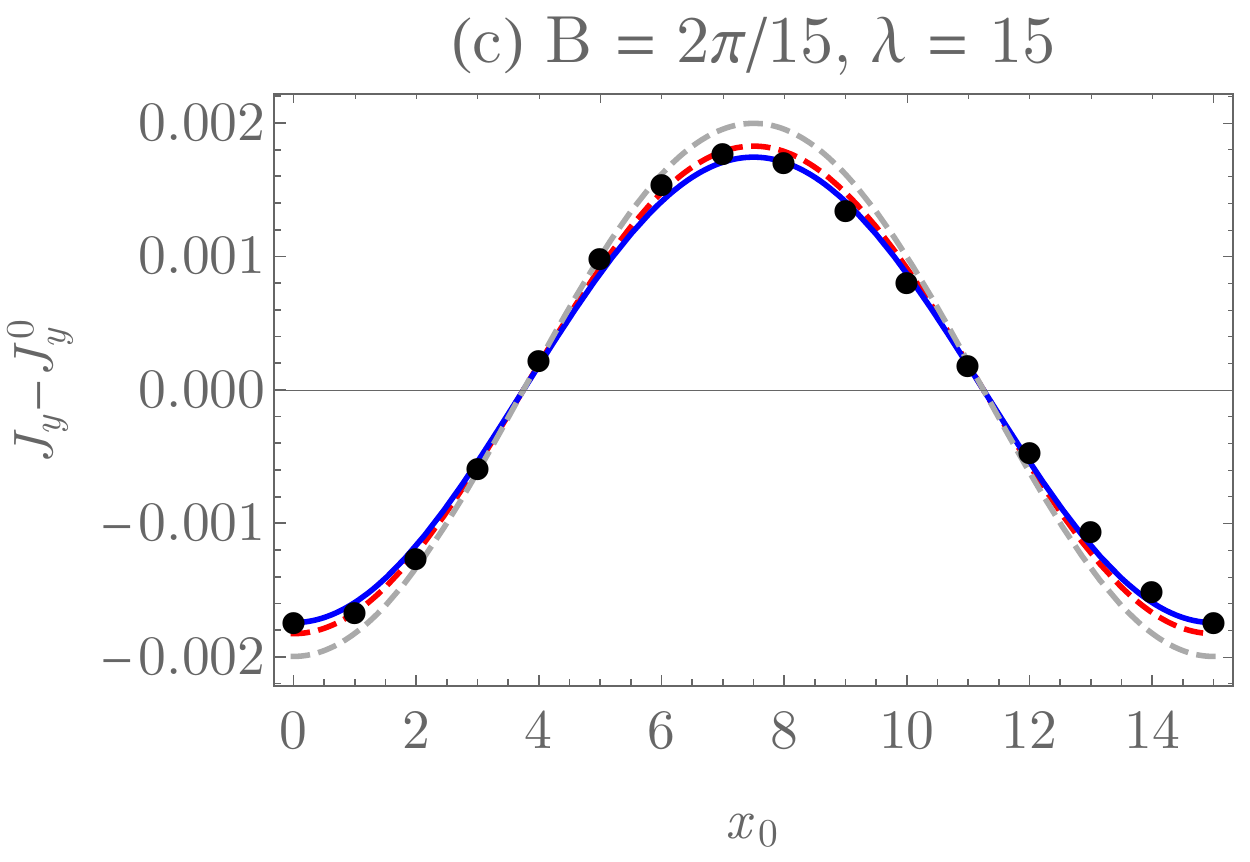}
\caption{Current density in the lowest band of the Hofstadter model in response to the applied sinusoidal electric potential of Eq.~\ref{eq:lattice_potential}, for (a) $B=2\pi/5$, $\lambda=5$; (b) $B=2\pi/10$, $\lambda=10$; and (c) $B=2\pi/15$, $\lambda=15$. Each plot shows the current density calculated numerically (black points), along with the analytic expression from Eq.~\eqref{eq:lattice_curr_dens_1} up to second derivatives without lattice corrections (grey dashed line); up to fourth derivatives without lattice corrections (red dashed line); and up to fourth derivatives with first order lattice corrections (solid blue line). The corrections from higher derivatives become less significant as the ratio $r=\lambda/l_B$ increases, which for the parameter sets above takes the values (a) $r=5.60$, (b) $r=7.93$ and (c) $r=9.71$. Leading lattice corrections are proportional to $1/N$. Exponentially small nonperturbative lattice corrections, corresponding to shorter wavelength oscillations, are also visible. In particular, the black numerical points oscillate with a small amplitude about the average (cosine) form (in blue) with half the wavelength, indicating leading nonperturbative corrections proportional to $\sin(2Nx_0)$. As motivated in Sec.~\ref{sec:perturbationtheory} (and expounded in Ref.~\cite{Harper:2014fq}), these exponentially small corrections arise due to tunnelling and lattice normalisation effects, and become negligible in the large $N$ limit.
\label{fig:hofstadter_numerics}}
\end{figure*}

Expanding the perturbation series consistently, we find that the current density for a $C_4$-symmetric lattice is, up to first order,
\begin{widetext}
\bequ
J_{yn'}^{C_4}(\br_0)=\left.J_{yn}(\br_0)+\frac{B}{2\pi}\bigg\{\frac{1}{24C^2_{2,0}}\bigg[6 \left(3 n^2+3 n+1\right) C^{4}_{2,2}+24 C^{4}_{4,0}\bigg]l_B^2\partial_x^2+\ldots\bigg\}E(x)\right|_{x=x_0},\label{eq:lattice_curr_dens_1}
\eequ
where $J_{yn}(\br_0)$ is the current density for the unperturbed $n$th Landau level given in Eq.~\eqref{eq:curr_dens_Ex}. In the above, we only give terms up to the second derivative of the electric field: further terms, and the specific expression for the Hofstadter model, may be found in Appendix~\ref{app:c4symmetry}.

As before, the leading homogeneous term is related to the quantised Hall conductance and receives no perturbative corrections. The higher-derivative terms, however, do receive corrections due to the lattice. Making the substitution $E(x)=E_qe^{iq\left(\hat{x}-x_0\right)}$ and using Eq.~\eqref{eq:j_sigma_e}, we identify the finite-wavevector Hall conductivity for the perturbed $n$th band as
\bequ
\sigma_{xyn'}(q)=\frac{1}{2\pi}\left\{1-\left[\frac{3}{2}\left(n+\frac{1}{2}\right)+\frac{B}{24C^2_{2,0}}\bigg(6 \left(3 n^2+3 n+1\right) C^{4}_{2,2}+24 C^{4}_{4,0}\bigg)\right]\frac{q^2}{B}+\ldots\right\},
\eequ
which has nonuniversal corrections at $O(Bq^2)$ (in comparison to Eq.~\eqref{eq:hall_cond_HS}). In an experiment, we would normally fill the lowest $K$ Landau levels or bands and measure their response simultaneously. The perturbed Hall conductivity for this situation (to be compared with Eq.~\eqref{eq:hall_cond_KLL}) can be obtained by summing the response above from bands $n=0$ to $n=K-1$, giving
\bequ
\sigma_{xy}^{K'}(q)=\frac{1}{2\pi}\left\{K-\left[\frac{3K^2}{4}+\frac{B}{4C^2_{2,0}}\bigg(C^4_{2,2}K^3+4C^4_{4,0}K\bigg)\right]\frac{q^2}{B}+\ldots\right\}.
\eequ
\end{widetext}
As for the isolated band, the $O(q^2)$ term, which one might measure to obtain the Hall viscosity, has nonuniversal corrections proportional to the magnetic field. These are a result of the rotational symmetry breaking due to the lattice, and would need to be accounted for in order to recover the appropriate continuum theory from an experimental measurement. In particular, if one naively applies the continuum expression from Eq.~\eqref{eq:HoyosSonResult} to a measured value of $\sigma_{xy}^{K'}(q)$, the extracted Hall viscosity $\tilde{\eta}_H$ would differ from the continuum value $\eta^{\rm cont}_H$ by the corrections
\beq
\frac{\tilde{\eta}_H}{\rho_0}&=&\frac{\eta^{\rm cont}_H}{\rho_0}-\frac{B}{4C^2_{2,0}}\bigg(C^4_{2,2}K^3+4C^4_{4,0}K\bigg).
\eeq
\subsection{Comparison with Numerics\label{sec:numerical_results}}
In order to verify this calculation, we compare Eq.~\eqref{eq:lattice_curr_dens_1} for the Hofstadter model (given explicitly in Appendix~\ref{app:c4symmetry}) with exact diagonalisation results for a range of parameter values. We choose an external sinusoidal potential of the form
\beq
V(x)&=&0.1\sin\left(\frac{2\pi x}{\lambda}\right)\label{eq:lattice_potential}
\eeq
and a magnetic field strength $B=2\pi/N$, for a range of $\lambda$ and $N$, and plot the exact current density in the lowest Hofstadter band in Fig.~\ref{fig:hofstadter_numerics}. Alongside this, we show the analytical expression from Eq.~\eqref{eq:lattice_curr_dens_1}, including up to: (i) terms proportional to $\partial_x^2E(x)$ with no lattice corrections; (ii) terms proportional to $\partial_x^4E(x)$ with no lattice corrections; and (iii) terms proportional to $\partial_x^4E(x)$ with lattice corrections at $O(B)$. In each of these plots, we have subtracted the zeroth order (DC) current density component,
\beq
J_y^0(\br_0)&=&\frac{1}{2\pi}E(x_0),
\eeq
so that the deviations can be seen more easily. We take the hopping parameter $t=N/4\pi$ so that the gap between low-lying bands is approximately $\Delta E=1$ and the relative strength of the external potential is approximately $V/\Delta E\approx 0.1$. 

Even though lattice corrections in the Hofstadter model are known to be fairly small \cite{Harper:2014fq}, their effects on the current density are clearly noticeable in Fig.~\ref{fig:hofstadter_numerics}. As found in the previous section, the leading corrections are proportional to $B=2\pi/N$ and so are most significant at small $N$. For the range of $N$ considered in Fig.~\ref{fig:hofstadter_numerics}, the lattice corrections calculated numerically show good agreement with our perturbative result. In addition to these, nonperturbative corrections (notably those proportional to $\sin(2Nx_0)$) are also visible in the numerical results in Fig.~\ref{fig:hofstadter_numerics}. 

The inclusion of higher derivative terms in Eq.~\eqref{eq:lattice_curr_dens_1}, which may be easily calculated using our formalism, also improves the agreement with the exact numerical results, as may be seen in Fig.~\ref{fig:hofstadter_numerics}. These corrections become less significant as the ratio of the external potential to the magnetic length, $\lambda/l_B$ gets larger. For a fixed $\lambda$ and $B$, more accurate approximations can be obtained by including higher terms in the double perturbation series of Eq.~\eqref{eq:lattice_curr_dens_1}.

\section{Conclusions\label{sec:conclusions}}
In this work, we have presented an approach for calculating the finite-wavevector electromagnetic response in lattice quantum Hall systems, illustrating our method with the explicit, relevant example of a lattice with $C_4$ rotational symmetry. In the process, we introduced a new, quantum mechanical derivation of the current response in quantum Hall systems, which allows terms to arbitrary order in the wavevector expansion to be calculated straightforwardly. We also developed a formalism for deriving the Landau-level-like wavefunctions that arise generically in tight-binding models in the presence of a magnetic field.

Our results have several implications for experiments that seek to measure the Hall viscosity through the inhomogeneous current response. Namely, at moderate field strengths, the lattice corrections to the current density can be significant, and must be taken into account if one wishes to extract the universal, Hall viscosity-dependent component that may be indicative of the underlying phase. Higher order derivative terms (quartic order in the wavevector and higher) may also be significant, but can be calculated straightforwardly within our formalism. The size of the corrections (and the field regime in which they are most noticeable) may depend sensitively on the structure of the underlying lattice model. Even in the Hofstadter model, for which lattice corrections are known to be fairly small \cite{Harper:2014fq}, corrections to the current response were found to be significant, and demonstrated good agreement between numerics and theory.

Beyond this, our perturbative approach to generic lattice models provides a universal framework for studying lattice quantum Hall systems, and is applicable to many different regimes of experimental interest. In many cases, only the band dispersion of the zero-field system is required to understand the finite-field behaviour. We demonstrated that the band structure endowed by a lattice generally leads to an effective continuum Hamiltonian with broken rotational symmetry. In this respect, such systems offer a natural realisation of the rotational symmetry-breaking Landau levels considered in Ref.~\cite{Haldane:2015tx}. In addition, we found that single-particle quantities, such as the current per orbital, are directly related to the band geometry of the system. If such a quantity can be detected, this would offer a means for measuring the band geometry experimentally. 

Our work raises a number of intriguing open questions. In addition to applying the approach to other lattice systems and real experiments, it would be of interest to study lattice corrections to the relativistic quantum Hall effect, which may arise in systems with Dirac-like dispersions such as graphene. Finally, some of the most interesting quantum Hall physics lies in the interacting regime. We leave a study of the electromagnetic response of such \emph{fractional} lattice quantum Hall systems to future work.

\begin{acknowledgements}
We are grateful to B.~Bradlyn, A.~Gromov, S.~H.~Simon, and A.~N.~Brown for useful discussions. The authors acknowledge support from the NSF under CAREER DMR-1455368 and the Alfred P. Sloan foundation. 
\end{acknowledgements}
\appendix
\section{Current Density from Linear Response Theory\label{app:j_linear_response}}
In this appendix, we show that the quantum mechanical approach to the calculation of current density used in Sec.~\ref{sec:QM_current_response} is equivalent to the more usual approach using linear response theory. We begin with the many-body Landau Hamiltonian, which we write in second quantisation as
\beq
\h_0&=&\sum_n\sum_{k_y}B\left[n+\frac{1}{2}\right]c^\dagger_{n,k_y}c^\ph_{n,k_y},
\eeq
where $c^\dagger_{n,k_y}$ creates a fermion in state $\ket{n,k_y}$. As before, we include the effect of a weak electric field by adding a static potential to the Hamiltonian. In many-body notation, this perturbing term can be written
\beq
\Delta \h&=&\int\dd^2\br\, V(x)c^\dagger (\br) c(\br),
\eeq
where $c^\dagger(\br)$ creates a fermion at position $\br$, and may be expressed alternatively in terms of Landau level operators as
\beq
c^\dagger(\br)&=&\sum_{n,k_y}\psi^*_{n,k_y}(\br)c^\dagger_{n,k_y}.
\eeq
The current density operator at position $\br_0$ takes the form
\beq
\hat{j}_y(\br_0)&=&c^\dagger(\br_0)\left(\hat{p}_y-Bx_0\right)c(\br_0).
\eeq

We now set up the system according to standard linear response techniques \cite{Mahan:2000wd,Bradlyn:2015vw}. Note, however, that we are representing the electric field as the spatial derivative of a scalar potential $V(x)$, rather than as the time derivative of a vector potential $\bA(t)$ as is more standard \footnote{See Appendix~A of Ref.~\cite{Chen:1989xs} for an example of the latter approach in a slightly different context.}. See Ref.~\cite{Bradlyn:2015vw} for a discussion of some of the subtle differences between the two choices.

We prepare the system, at $t=-\infty$, in an eigenstate of the Hamiltonian $\h_0$, in this case choosing the filled $n$th Landau level, 
\beq
\ket{nLL}&=&\prod_{k_y}c^\dagger_{n,k_y}\ket{0}.\label{eq:nll}
\eeq
We then adiabatically switch on the static potential. To facilitate this, we introduce a time dependence to $V(x)$ through
\be
V(x)\to V(x,t)&=&V(x)e^{-i\omega_+ t},
\ee
where $\omega_+=\omega +i\epsilon$. In this way, the perturbing potential vanishes at $t=-\infty$, and at $t=0$, the system is described by the complete Hamiltonian $\h=\h_0+\Delta \h$. At the end of the calculation we will set $\omega_+\to0$.

We work in the interaction picture, where operators and states gain time dependence (relative to the Schr\"odinger picture) through
\beq
\hat{O}_I(t)&=&e^{i\h_0t}\hat{O}e^{-i\h_0t},\\
\ket{\psi(t)}_I&=&e^{i\h_0t}\ket{\psi}.
\eeq
States evolve according to the time evolution 
\beq
\ket{\psi(t)}_I&=&U(t,t_0)\ket{\psi(t_0)}_I
\eeq
with
\beq
U(t,t_0)&=&\mathcal{T}\exp\left[-i\int_{t_0}^t\Delta\h_I(t')\dd t'\right],
\eeq
where $\mathcal{T}$ is the time ordering operator. The unperturbed state $\ket{nLL}$ at $t=-\infty$ evolves to the perturbed, interaction-picture state $\ket{nLL(t)}_I$ at time $t$ through
\beq
\ket{nLL(t)}_I&=&U(t,-\infty)\ket{nLL}.
\eeq
We can then find the current density by calculating the expectation value $_I\!\bra{nLL(t)}\hat{j}_{Iy}(\br_0,t)\ket{nLL(t)}_I$. 

Expanding the time-evolution operator to first order in the perturbation, we obtain the linear response result
\begin{widetext}
\beq
\left\langle j_{Iyn}(\br_0,t)\right\rangle&=&\bra{nLL}j_{Iy}(\br_0,t)\ket{nLL}+i\int_{-\infty}^t\dd t'\bra{nLL}\left[\Delta H_I(t'),j_{Iy}(\br_0,t)\right]\ket{nLL},\label{eq:jiy_lin_resp}
\eeq
\end{widetext}
where the expectation values are now taken with respect to the unperturbed initial state. To simplify notation, we write
\bequ
\Delta \h_I(t)=e^{-i\omega_+ t}e^{iH_0t}\sum_{\{n\},\{k\}}V^{n_1,k_1}_{n_2,k_2}c^\dagger_{n_1,k_1}c^\ph_{n_2,k_2}e^{-iH_0t}
\eequ
with 
\beq
V^{n_1,k_1}_{n_2,k_2}&=&\int\dd^2 \br \,\psi^*_{n_1,k_1}(\br)V(x)\psi_{n_2,k_2}(\br)\nonumber\\
&\equiv&V^{n_1,k_1}_{n_2,k_1}\delta_{k1,k_2}\label{eq:dH_simple}
\eeq
and
\bequ
j_{Iy}(\br_0,t)=e^{iH_0t}\sum_{\{n\},\{k\}}j^{n_3,k_3}_{n_4,k_4}(\br_0)c^\dagger_{n_3,k_3}c^\ph_{n_4,k_4}e^{-iH_0t}\label{eq:jr_simple}
\eequ
with
\beq
j^{n_3,k_3}_{n_4,k_4}(\br_0)&=&\psi_{n_3,k_3}^*(\br_0) \hat{v}_y\psi_{n_4,k_4}(\br_0).
\eeq
Then, inserting Eqs.~\eqref{eq:nll}, \eqref{eq:dH_simple} and \eqref{eq:jr_simple} into Eq.~\eqref{eq:jiy_lin_resp}, manipulating the field operators, and integrating over $t'$, we obtain
\begin{widetext}
\bequ
\langle j_{Iyn}(\br_0,t)\rangle=-e^{-i\omega_+ t}\sum_{m\neq n,k}\frac{V^{n,k}_{m,k}j^{m,k}_{n,k}(\br_0)}{\omega+i\epsilon+E_m-E_n}+e^{-i\omega_+ t}\sum_{m\neq n,k}\frac{V^{m,k}_{n,k}j^{n,k}_{m,k}(\br_0)}{\omega+i\epsilon-E_m+E_n},
\eequ
\end{widetext}
where several contributions have cancelled out. Since $E_m\neq E_n$, we may now take the $\omega_+\to0$ limit directly to find
\bequ
\langle j_{Iyn}(q,\omega\to0)\rangle=\sum_{m\neq n,k}\frac{V^{n,k}_{m,k}j^{m,k}_{n,k}(\br_0)+V^{m,k}_{n,k}j^{n,k}_{m,k}(\br_0)}{E_n-E_m}.
\eequ
By substituting for $V^{n,k}_{m,k}$ and $j^{m,k}_{n,k}(\br_0)$, we recover Eq.~\eqref{eq:curr_dens_1}, obtained in Sec.~\ref{sec:QM_current_response} using more elementary means.
\section{Perturbative Results for Bravais Lattices with $C_4$ Symmetry\label{app:c4symmetry}}
In this Appendix, we give some explicit perturbative results for tight-binding models on a generic two-dimensional lattice with $C_4$ symmetry. We also give some specific results for the Hofstadter model \cite{Harper:1955uu,Hofstadter:1976wt}.
\subsection{Wavefunctions and Energies}
The generic zero-field band structure for a lattice with $C_4$ symmetry was given in Eq.~\eqref{eq:C4bandstructure}. In the presence of a weak magnetic field, the substitution $\bk\to\hat{\bk}$ may be enacted to obtain Eqs.~\eqref{eq:C4Ham1} and \eqref{eq:C4Ham2}. In terms of ladder operators (defined in Eq.~\eqref{eq:k_ladders}), this Hamiltonian may be written
\beq
\h_{C_4}^{(1)}&=&C_{0,0}^0+2BC^2_{2,0}\left(a^\dagger a+\frac{1}{2}\right)\\
\h_{C_4}^{(2)}&=&\frac{C^4_{4,0}B^2}{2}\left[3+\an{4}+\ad{4}+12a^\dagger a+6\ad{2}\an{2}\right]\nonumber\\
&&+\frac{iC^4_{3,1}B^2}{2}\left[\ad{4}-\an{4}\right]\\
&&-\frac{C^4_{2,2}B^2}{4}\left[-1+\an{4}+\ad{4}-4a^\dagger a-2\ad{2}\an{2}\right].\nonumber
\eeq
Higher order terms may be found similarly. 

At lowest order, we solve $\h_{C_4}^{(1)}$ to find Landau level-like solutions. At next order, we treat $\h_{C_4}^{(2)}$ at first order in perturbation theory to obtain
\beq
E_{C_4}^{(1,2)}(n)&=&C^0_{0,0}+2BC^2_{2,0}\left(n+\frac{1}{2}\right)\\
&&+\frac{3C^4_{4,0}B^2}{2}\left(1+2n+2n^2\right)\nonumber\\
&&+\frac{C^4_{2,2}B^2}{4}\left(1+2n+2n^2\right)\nonumber
\eeq
and
\beq
\ket{n}^{(1,2)}_{C_4}&=&\ket{n}+\frac{B}{32C^2_{2,0}}\left[2C^4_{4,0}-2iC^4_{3,1}-C^4_{2,2}\right]\an{4}\ket{n}\nonumber\\
&&-\frac{B}{32C^2_{2,0}}\left[2C^4_{4,0}+2iC^4_{3,1}-C^4_{2,2}\right]\ad{4}\ket{n}\nonumber\\
&&
\eeq
where $\ket{n}$ are unperturbed Landau levels. The perturbed states may be obtained from the unperturbed states through the action of a unitary operator, $\ket{n}^{(1,2)}_{C_4}=U^\dagger\ket{n}$, with
\beq
U^\dagger&=&\exp\bigg[\frac{B}{32C^2_{2,0}}\left[2C^4_{4,0}-C^4_{2,2}\right]\left[\an{4}-\ad{4}\right]\nonumber\\
&&-\frac{2iBC^4_{3,1}}{32C^2_{2,0}}\left[\an{4}+\ad{4}\right]\bigg]
\eeq
to leading order.

The Hofstadter model is a particular example of a lattice with $C_4$ symmetry, with coefficients (setting $t=1$) $C^0_{0,0}=-4$, $C^2_{2,0}=1$, $C^4_{4,0}=-1/12$ and $C^4_{3,1}=C^4_{2,2}=0$. With these substitutions, the perturbed energies and wavefunctions are
\beq
E^{(1,2)}_{\rm Hof}(n)&=&-4+2B\left(n+\frac{1}{2}\right)-\frac{B^2}{8}\left(1+2n+2n^2\right)\nonumber\\
\ket{n}^{(1,2)}_{\rm Hof}&=&\ket{n}-\frac{B}{192}\an{4}\ket{n}+\frac{B}{192}\ad{4}\ket{n},
\eeq
and the unitary operator $U^\dagger$ takes the form
\beq
U^\dagger&=&\exp\bigg[-\frac{B}{192}\left[\an{4}-\ad{4}\right]\bigg].
\eeq
Higher order results are given in Ref.~\cite{Harper:2014fq}.
\subsection{Current Response}
In Sec.~\ref{sec:em_lattice} we defined the perturbed current operator that arises as a result of the lattice. For $C_4$ symmetric lattices, this operator is given in terms of $\kx$ and $\ky$ in Eq.~\eqref{eq:lat_current_momentum}. This may be written more usefully in terms of ladder operators as
\beq
&&\hat{I}^{C_4}_{y}=-\sqrt{2B}C^2_{2,0}\left(a+a^\dagger\right)-\frac{B^{3/2}}{\sqrt{2}}\bigg\{\left(6C^4_{4,0}+C^4_{2,2}\right)\left(a^\dagger+a\right)\nonumber\\
&&+\left(2C^4_{4,0}-C^4_{2,2}\right)\left(\ad{3}+\an{3}\right)+2iC^4_{3,1}\left(\ad{3}-\an{3}\right)\nonumber\\
&&+\left(6C^4_{4,0}+C^4_{2,2}\right)\left(a^\dagger\an{2}+\ad{2}a\right).
\eeq
For the Hofstadter model, this operator takes the form
\beq
\hat{I}^{\mathrm{Hof}}_{y}&=&-\sqrt{2B}\left(a+\ado\right)+\left(\frac{B}{2}\right)^{3/2}\bigg[\frac{1}{3}\ad{3}+\frac{1}{3}\an{3}\nonumber\\
&&+a+\ado+\ad{2}a+\ado\an{2}\bigg].
\eeq

In Sec.~\ref{sec:lattice_curr_orb}, we demonstrated how to calculate the expectation value of this operator for a single particle state in a lattice with $C_4$ symmetry, resulting in the current per orbital given in Eq.~\eqref{eq:curr_orb_c4}. For the specific case of the Hofstadter model, this expression yields, to first order in $B$,
\small
\begin{align}
\left\langle\hat{I}_{yn'}^{\rm Hof}\right\rangle=&l_B^2\left[1+\frac{1}{2}\left(n+\frac{1}{2}\right)l_B^2\partial_x^2\right.\\
&+\left\{\frac{1}{16}\left(n^2+n+\frac{1}{2}\right)\right.\nonumber\\
&\left.\left.\left.+B\frac{\left(2n+1\right)}{1152}\left(n^2+n+3\right)\right\}l_B^4\partial_x^4+\ldots\right]E(x)\right|_{x=\frac{k_y}{B}}.\nonumber
\end{align}
\normalsize
Higher order corrections due to the lattice and higher derivatives of the electric field may be included straightforwardly using the methods described in the main text.

In Sec.~\ref{sec:lattice_curr_dens}, we calculated the current density response for a lattice with $C_4$ symmetry, obtaining the result in Eq.~\eqref{eq:lattice_curr_dens_1}. Here, we give a more complete expression that includes terms involving the fourth derivative of the external field,
\begin{widetext}
\beq
J_{yn'}^{C_4}(\br_0)&=&J_{yn}(\br_0)+\frac{B}{2\pi}\bigg\{\frac{1}{24C^2_{2,0}}\bigg[6 \left(3 n^2+3 n+1\right) C^{4}_{2,2}+24 C^{4}_{4,0}\bigg]l_B^2\partial_x^2\nonumber\\
&&+\frac{1}{1152 C^{2}_{2,0}} \bigg[2 \left(190 n^3+285 n^2+197 n+51\right) C^{4}_{2,2}-60 \left(2 n^3+3 n^2-13 n-7\right) C^{4}_{4,0}\left.\bigg]l_B^4\partial_x^4+\ldots\bigg\}E(x)\right|_{x=x_0}.\nonumber
\eeq
For the specific case of the Hofstadter model, this becomes
\beq
J_{yn'}^{\rm Hof}(\br_0)&=&\frac{1}{2\pi}\left[1+\left\{\frac{3}{2}\left(n+\frac{1}{2}\right)-\frac{B}{12}\right\}l_B^2\partial_x^2\right.\label{eq:curr_dens_Hof}\\
&&\left.\left.+\left\{\frac{1}{48}\left(11+30n+30n^2\right)+\frac{5B}{1152}\left(2n^3+3n^2-13n-7\right)\right\}l_B^4\partial_x^4+\ldots\right]E(x)\right|_{x=x_0},\nonumber
\eeq
from which we can extract the Hall conductivity
\beq
\sigma_{xyn'}^{\rm Hof}(q)&=&\frac{1}{2\pi}\left[1-\left\{\frac{3}{2}\left(n+\frac{1}{2}\right)-\frac{B}{12}\right\}\frac{q^2}{B}\right.\label{eq:Hall_cond_Hof}\\
&&\left.\left.+\left\{\frac{1}{48}\left(11+30n+30n^2\right)+\frac{5B}{1152}\left(2n^3+3n^2-13n-7\right)\right\}\frac{q^4}{B^2}+\ldots\right]E(x)\right|_{x=x_0}.\nonumber
\eeq
If we fill the lowest $K$ bands, the combined response of the set of bands is the sum of the above from $n=0$ to $n=K-1$, which gives
\bequ
\sigma_{xy}^{{\rm Hof},K'}(q)=\left.\frac{1}{2\pi}\bigg[K-\left\{\frac{3K^2}{4}-B\frac{K}{12}\right\}\frac{q^2}{B}-\left\{\left(\frac{5K^3}{24}+\frac{K}{48}\right)+B\left(\frac{5K^4}{2304}-\frac{25K^2}{768}\right)\right\}\frac{q^4}{B^2}+\ldots\bigg]E(x)\right|_{x=x_0}.
\eequ
\end{widetext}


\begin{thebibliography}{56}%
\makeatletter
\providecommand \@ifxundefined [1]{%
 \@ifx{#1\undefined}
}%
\providecommand \@ifnum [1]{%
 \ifnum #1\expandafter \@firstoftwo
 \else \expandafter \@secondoftwo
 \fi
}%
\providecommand \@ifx [1]{%
 \ifx #1\expandafter \@firstoftwo
 \else \expandafter \@secondoftwo
 \fi
}%
\providecommand \natexlab [1]{#1}%
\providecommand \enquote  [1]{``#1''}%
\providecommand \bibnamefont  [1]{#1}%
\providecommand \bibfnamefont [1]{#1}%
\providecommand \citenamefont [1]{#1}%
\providecommand \href@noop [0]{\@secondoftwo}%
\providecommand \href [0]{\begingroup \@sanitize@url \@href}%
\providecommand \@href[1]{\@@startlink{#1}\@@href}%
\providecommand \@@href[1]{\endgroup#1\@@endlink}%
\providecommand \@sanitize@url [0]{\catcode `\\12\catcode `\$12\catcode
  `\&12\catcode `\#12\catcode `\^12\catcode `\_12\catcode `\%12\relax}%
\providecommand \@@startlink[1]{}%
\providecommand \@@endlink[0]{}%
\providecommand \url  [0]{\begingroup\@sanitize@url \@url }%
\providecommand \@url [1]{\endgroup\@href {#1}{\urlprefix }}%
\providecommand \urlprefix  [0]{URL }%
\providecommand \Eprint [0]{\href }%
\providecommand \doibase [0]{http://dx.doi.org/}%
\providecommand \selectlanguage [0]{\@gobble}%
\providecommand \bibinfo  [0]{\@secondoftwo}%
\providecommand \bibfield  [0]{\@secondoftwo}%
\providecommand \translation [1]{[#1]}%
\providecommand \BibitemOpen [0]{}%
\providecommand \bibitemStop [0]{}%
\providecommand \bibitemNoStop [0]{.\EOS\space}%
\providecommand \EOS [0]{\spacefactor3000\relax}%
\providecommand \BibitemShut  [1]{\csname bibitem#1\endcsname}%
\let\auto@bib@innerbib\@empty
\bibitem [{\citenamefont {Prange}\ and\ \citenamefont
  {Girvin}(1987)}]{prange1987quantum}%
  \BibitemOpen
  \bibfield  {author} {\bibinfo {author} {\bibfnamefont {R~E}\ \bibnamefont
  {Prange}}\ and\ \bibinfo {author} {\bibfnamefont {S~M}\ \bibnamefont
  {Girvin}},\ }\href {http://books.google.co.uk/books?id=Y7XvAAAAMAAJ} {\emph
  {\bibinfo {title} {{The Quantum Hall Effect}}}},\ Graduate texts in
  contemporary physics\ (\bibinfo  {publisher} {Springer-Verlag},\ \bibinfo
  {address} {New York},\ \bibinfo {year} {1987})\BibitemShut {NoStop}%
\bibitem [{\citenamefont {Klitzing}\ \emph {et~al.}(1980)\citenamefont
  {Klitzing}, \citenamefont {Dorda},\ and\ \citenamefont
  {Pepper}}]{Klitzing:428079}%
  \BibitemOpen
  \bibfield  {author} {\bibinfo {author} {\bibfnamefont {K~V}\ \bibnamefont
  {Klitzing}}, \bibinfo {author} {\bibfnamefont {G}~\bibnamefont {Dorda}}, \
  and\ \bibinfo {author} {\bibfnamefont {M}~\bibnamefont {Pepper}},\ }\bibfield
   {title} {\enquote {\bibinfo {title} {{New method for high-accuracy
  determination of the fine-structure constant based on quantized Hall
  resistance}},}\ }\href@noop {} {\bibfield  {journal} {\bibinfo  {journal}
  {Physical Review Letters}\ }\textbf {\bibinfo {volume} {45}},\ \bibinfo
  {pages} {494--497} (\bibinfo {year} {1980})}\BibitemShut {NoStop}%
\bibitem [{\citenamefont {Tsui}\ \emph {et~al.}(1982)\citenamefont {Tsui},
  \citenamefont {Stormer},\ and\ \citenamefont {Gossard}}]{Tsui:1982yy}%
  \BibitemOpen
  \bibfield  {author} {\bibinfo {author} {\bibfnamefont {D~C}\ \bibnamefont
  {Tsui}}, \bibinfo {author} {\bibfnamefont {H~L}\ \bibnamefont {Stormer}}, \
  and\ \bibinfo {author} {\bibfnamefont {A~C}\ \bibnamefont {Gossard}},\
  }\bibfield  {title} {\enquote {\bibinfo {title} {{Two-dimensional
  magnetotransport in the extreme quantum limit}},}\ }\href {\doibase
  10.1103/PhysRevLett.48.1559} {\bibfield  {journal} {\bibinfo  {journal}
  {Physical Review Letters}\ }\textbf {\bibinfo {volume} {48}},\ \bibinfo
  {pages} {1559--1562} (\bibinfo {year} {1982})}\BibitemShut {NoStop}%
\bibitem [{\citenamefont {Lifshit︠z}\ and\ \citenamefont
  {Pitaevskii}(1981)}]{Lifshitz:2012th}%
  \BibitemOpen
  \bibfield  {author} {\bibinfo {author} {\bibfnamefont {E~M}\ \bibnamefont
  {Lifshit︠z}}\ and\ \bibinfo {author} {\bibfnamefont {L~P}\ \bibnamefont
  {Pitaevskii}},\ }\href
  {http://books.google.com/books?id=DTHxPDfV0fQC&printsec=frontcover&dq=intitle:Physical+Kinetics&hl=&cd=1&source=gbs_api}
  {\emph {\bibinfo {title} {{Landau and Lifshitz, Course of Theoretical
  Physics}}}},\ Volume 10: Physical Kinetics\ (\bibinfo  {publisher}
  {Butterworth-Heinemann},\ \bibinfo {year} {1981})\BibitemShut {NoStop}%
\bibitem [{\citenamefont {L{\'e}vay}(1995)}]{Levay:1995dq}%
  \BibitemOpen
  \bibfield  {author} {\bibinfo {author} {\bibfnamefont {P{\'e}ter}\
  \bibnamefont {L{\'e}vay}},\ }\bibfield  {title} {\enquote {\bibinfo {title}
  {{Berry phases for Landau Hamiltonians on deformed tori}},}\ }\href {\doibase
  10.1063/1.531066} {\bibfield  {journal} {\bibinfo  {journal} {Journal of
  Mathematical Physics}\ }\textbf {\bibinfo {volume} {36}},\ \bibinfo {pages}
  {2792} (\bibinfo {year} {1995})}\BibitemShut {NoStop}%
\bibitem [{\citenamefont {Avron}\ \emph {et~al.}(1995)\citenamefont {Avron},
  \citenamefont {Seiler},\ and\ \citenamefont {Zograf}}]{Avron:276305}%
  \BibitemOpen
  \bibfield  {author} {\bibinfo {author} {\bibfnamefont {J~E}\ \bibnamefont
  {Avron}}, \bibinfo {author} {\bibfnamefont {R}~\bibnamefont {Seiler}}, \ and\
  \bibinfo {author} {\bibfnamefont {P~G}\ \bibnamefont {Zograf}},\ }\bibfield
  {title} {\enquote {\bibinfo {title} {{Viscosity of quantum Hall fluids}},}\
  }\href {http://cds.cern.ch/record/276305} {\bibfield  {journal} {\bibinfo
  {journal} {Physical Review Letters}\ }\textbf {\bibinfo {volume} {75}},\
  \bibinfo {pages} {697--700. 14 p} (\bibinfo {year} {1995})}\BibitemShut
  {NoStop}%
\bibitem [{\citenamefont {Read}(2009)}]{Read:2009iw}%
  \BibitemOpen
  \bibfield  {author} {\bibinfo {author} {\bibfnamefont {N}~\bibnamefont
  {Read}},\ }\bibfield  {title} {\enquote {\bibinfo {title} {{Non-Abelian
  adiabatic statistics and Hall viscosity in quantum Hall states and
  $p_{x}+ip_{y}$ paired superfluids}},}\ }\href {\doibase
  10.1103/PhysRevB.79.045308} {\bibfield  {journal} {\bibinfo  {journal}
  {Physical Review B}\ }\textbf {\bibinfo {volume} {79}},\ \bibinfo {pages}
  {045308} (\bibinfo {year} {2009})}\BibitemShut {NoStop}%
\bibitem [{\citenamefont {Read}\ and\ \citenamefont
  {Rezayi}(2011)}]{Read:2011hx}%
  \BibitemOpen
  \bibfield  {author} {\bibinfo {author} {\bibfnamefont {N}~\bibnamefont
  {Read}}\ and\ \bibinfo {author} {\bibfnamefont {E~H}\ \bibnamefont
  {Rezayi}},\ }\bibfield  {title} {\enquote {\bibinfo {title} {{Hall viscosity,
  orbital spin, and geometry: Paired superfluids and quantum Hall systems}},}\
  }\href {\doibase 10.1103/PhysRevB.84.085316} {\bibfield  {journal} {\bibinfo
  {journal} {Physical Review B}\ }\textbf {\bibinfo {volume} {84}},\ \bibinfo
  {pages} {085316} (\bibinfo {year} {2011})}\BibitemShut {NoStop}%
\bibitem [{\citenamefont {Wen}\ and\ \citenamefont {Zee}(1992)}]{Wen:1992ug}%
  \BibitemOpen
  \bibfield  {author} {\bibinfo {author} {\bibfnamefont {Xiao-Gang}\
  \bibnamefont {Wen}}\ and\ \bibinfo {author} {\bibfnamefont {A}~\bibnamefont
  {Zee}},\ }\bibfield  {title} {\enquote {\bibinfo {title} {{Shift and Spin
  Vector - New Topological Quantum Numbers for the Hall Fluids}},}\ }\href
  {http://journals.aps.org/prl/abstract/10.1103/PhysRevLett.69.953} {\bibfield
  {journal} {\bibinfo  {journal} {Physical Review Letters}\ }\textbf {\bibinfo
  {volume} {69}},\ \bibinfo {pages} {953--956} (\bibinfo {year}
  {1992})}\BibitemShut {NoStop}%
\bibitem [{\citenamefont {Hoyos}\ and\ \citenamefont
  {Son}(2012)}]{Hoyos:2012eu}%
  \BibitemOpen
  \bibfield  {author} {\bibinfo {author} {\bibfnamefont {Carlos}\ \bibnamefont
  {Hoyos}}\ and\ \bibinfo {author} {\bibfnamefont {Dam~Thanh}\ \bibnamefont
  {Son}},\ }\bibfield  {title} {\enquote {\bibinfo {title} {{Hall Viscosity and
  Electromagnetic Response}},}\ }\href {\doibase
  10.1103/PhysRevLett.108.066805} {\bibfield  {journal} {\bibinfo  {journal}
  {Physical Review Letters}\ }\textbf {\bibinfo {volume} {108}},\ \bibinfo
  {pages} {066805} (\bibinfo {year} {2012})}\BibitemShut {NoStop}%
\bibitem [{\citenamefont {Bradlyn}\ \emph {et~al.}(2012)\citenamefont
  {Bradlyn}, \citenamefont {Goldstein},\ and\ \citenamefont
  {Read}}]{Bradlyn:2012he}%
  \BibitemOpen
  \bibfield  {author} {\bibinfo {author} {\bibfnamefont {Barry}\ \bibnamefont
  {Bradlyn}}, \bibinfo {author} {\bibfnamefont {Moshe}\ \bibnamefont
  {Goldstein}}, \ and\ \bibinfo {author} {\bibfnamefont {N}~\bibnamefont
  {Read}},\ }\bibfield  {title} {\enquote {\bibinfo {title} {{Kubo formulas for
  viscosity: Hall viscosity, Ward identities, and the relation with
  conductivity}},}\ }\href {\doibase 10.1103/PhysRevB.86.245309} {\bibfield
  {journal} {\bibinfo  {journal} {Physical Review B}\ }\textbf {\bibinfo
  {volume} {86}},\ \bibinfo {pages} {245309} (\bibinfo {year}
  {2012})}\BibitemShut {NoStop}%
\bibitem [{\citenamefont {Biswas}(2013)}]{Biswas:2013wm}%
  \BibitemOpen
  \bibfield  {author} {\bibinfo {author} {\bibfnamefont {Rudro~R}\ \bibnamefont
  {Biswas}},\ }\bibfield  {title} {\enquote {\bibinfo {title} {{Semiclassical
  theory of viscosity in quantum Hall states}},}\ }\href
  {http://arxiv.org/abs/1311.7149v2} {\bibfield  {journal} {\bibinfo  {journal}
  {arXiv}\ } (\bibinfo {year} {2013})},\ \Eprint
  {http://arxiv.org/abs/1311.7149v2} {1311.7149v2} \BibitemShut {NoStop}%
\bibitem [{\citenamefont {Haldane}\ and\ \citenamefont
  {Shen}(2015)}]{Haldane:2015tx}%
  \BibitemOpen
  \bibfield  {author} {\bibinfo {author} {\bibfnamefont {F~D~M}\ \bibnamefont
  {Haldane}}\ and\ \bibinfo {author} {\bibfnamefont {Yu}~\bibnamefont {Shen}},\
  }\bibfield  {title} {\enquote {\bibinfo {title} {{Geometry of Landau orbits
  in the absence of rotational symmetry}},}\ }\href
  {http://arxiv.org/abs/1512.04502v2} {\bibfield  {journal} {\bibinfo
  {journal} {arXiv}\ } (\bibinfo {year} {2015})},\ \Eprint
  {http://arxiv.org/abs/1512.04502v2} {1512.04502v2} \BibitemShut {NoStop}%
\bibitem [{\citenamefont {Huang}(2015)}]{Huang:2015ga}%
  \BibitemOpen
  \bibfield  {author} {\bibinfo {author} {\bibfnamefont {Biao}\ \bibnamefont
  {Huang}},\ }\bibfield  {title} {\enquote {\bibinfo {title} {{Hall viscosity
  revealed via density response}},}\ }\href {\doibase
  10.1103/PhysRevB.91.235101} {\bibfield  {journal} {\bibinfo  {journal}
  {Physical Review B}\ }\textbf {\bibinfo {volume} {91}},\ \bibinfo {pages}
  {235101--5} (\bibinfo {year} {2015})}\BibitemShut {NoStop}%
\bibitem [{\citenamefont {Scaffidi}\ \emph {et~al.}(2017)\citenamefont
  {Scaffidi}, \citenamefont {Nandi}, \citenamefont {Schmidt}, \citenamefont
  {Mackenzie},\ and\ \citenamefont {Moore}}]{Scaffidi:2017fb}%
  \BibitemOpen
  \bibfield  {author} {\bibinfo {author} {\bibfnamefont {Thomas}\ \bibnamefont
  {Scaffidi}}, \bibinfo {author} {\bibfnamefont {Nabhanila}\ \bibnamefont
  {Nandi}}, \bibinfo {author} {\bibfnamefont {Burkhard}\ \bibnamefont
  {Schmidt}}, \bibinfo {author} {\bibfnamefont {Andrew~P}\ \bibnamefont
  {Mackenzie}}, \ and\ \bibinfo {author} {\bibfnamefont {Joel~E}\ \bibnamefont
  {Moore}},\ }\bibfield  {title} {\enquote {\bibinfo {title} {{Hydrodynamic
  Electron Flow and Hall Viscosity}},}\ }\href {\doibase
  10.1103/PhysRevLett.118.226601} {\bibfield  {journal} {\bibinfo  {journal}
  {Physical Review Letters}\ }\textbf {\bibinfo {volume} {118}},\ \bibinfo
  {pages} {771--5} (\bibinfo {year} {2017})}\BibitemShut {NoStop}%
\bibitem [{\citenamefont {Pellegrino}\ \emph {et~al.}(2017)\citenamefont
  {Pellegrino}, \citenamefont {Torre},\ and\ \citenamefont
  {Polini}}]{Pellegrino:2017fd}%
  \BibitemOpen
  \bibfield  {author} {\bibinfo {author} {\bibfnamefont {Francesco M~D}\
  \bibnamefont {Pellegrino}}, \bibinfo {author} {\bibfnamefont {Iacopo}\
  \bibnamefont {Torre}}, \ and\ \bibinfo {author} {\bibfnamefont {Marco}\
  \bibnamefont {Polini}},\ }\bibfield  {title} {{\selectlanguage
  {English}\enquote {\bibinfo {title} {{Nonlocal transport and the Hall
  viscosity of two-dimensional hydrodynamic electron liquids}},}\ }}\href
  {\doibase 10.1103/PhysRevB.96.195401} {\bibfield  {journal} {\bibinfo
  {journal} {Physical Review B}\ }\textbf {\bibinfo {volume} {96}},\ \bibinfo
  {pages} {195401--11} (\bibinfo {year} {2017})}\BibitemShut {NoStop}%
\bibitem [{\citenamefont {Delacretaz}\ and\ \citenamefont
  {Gromov}(2017)}]{Delacretaz:2017kx}%
  \BibitemOpen
  \bibfield  {author} {\bibinfo {author} {\bibfnamefont {Luca~V}\ \bibnamefont
  {Delacretaz}}\ and\ \bibinfo {author} {\bibfnamefont {Andrey}\ \bibnamefont
  {Gromov}},\ }\bibfield  {title} {{\selectlanguage {English}\enquote {\bibinfo
  {title} {{Transport Signatures of the Hall Viscosity}},}\ }}\href {\doibase
  10.1103/PhysRevLett.119.226602} {\bibfield  {journal} {\bibinfo  {journal}
  {Physical Review Letters}\ }\textbf {\bibinfo {volume} {119}},\ \bibinfo
  {pages} {226602--5} (\bibinfo {year} {2017})}\BibitemShut {NoStop}%
\bibitem [{\citenamefont {Gromov}\ and\ \citenamefont
  {Son}(2017)}]{Gromov:2017io}%
  \BibitemOpen
  \bibfield  {author} {\bibinfo {author} {\bibfnamefont {Andrey}\ \bibnamefont
  {Gromov}}\ and\ \bibinfo {author} {\bibfnamefont {Dam~Thanh}\ \bibnamefont
  {Son}},\ }\bibfield  {title} {{\selectlanguage {English}\enquote {\bibinfo
  {title} {{Bimetric Theory of Fractional Quantum Hall States}},}\ }}\href
  {\doibase 10.1103/PhysRevX.7.041032} {\bibfield  {journal} {\bibinfo
  {journal} {Physical Review X}\ }\textbf {\bibinfo {volume} {7}},\ \bibinfo
  {pages} {041032--15} (\bibinfo {year} {2017})}\BibitemShut {NoStop}%
\bibitem [{\citenamefont {Contreras}\ \emph {et~al.}(1997)\citenamefont
  {Contreras}, \citenamefont {Knap},\ and\ \citenamefont
  {Skierbiszewski}}]{Contreras:1997uc}%
  \BibitemOpen
  \bibfield  {author} {\bibinfo {author} {\bibfnamefont {S}~\bibnamefont
  {Contreras}}, \bibinfo {author} {\bibfnamefont {W}~\bibnamefont {Knap}}, \
  and\ \bibinfo {author} {\bibfnamefont {C}~\bibnamefont {Skierbiszewski}},\
  }\bibfield  {title} {\enquote {\bibinfo {title} {{Observation of quantum Hall
  effect in 2D-electron gas confined in GaN/GaAlN heterostructure}},}\ }\href
  {http://www.sciencedirect.com/science/article/pii/S0921510796019393}
  {\bibfield  {journal} {\bibinfo  {journal} {Mat. Sci. Eng. B}\ }\textbf
  {\bibinfo {volume} {46}},\ \bibinfo {pages} {92} (\bibinfo {year}
  {1997})}\BibitemShut {NoStop}%
\bibitem [{\citenamefont {Tsukazaki}\ \emph {et~al.}(2007)\citenamefont
  {Tsukazaki}, \citenamefont {Ohtomo}, \citenamefont {Kita}, \citenamefont
  {Ohno}, \citenamefont {Ohno},\ and\ \citenamefont
  {Kawasaki}}]{Tsukazaki:2007be}%
  \BibitemOpen
  \bibfield  {author} {\bibinfo {author} {\bibfnamefont {A}~\bibnamefont
  {Tsukazaki}}, \bibinfo {author} {\bibfnamefont {A}~\bibnamefont {Ohtomo}},
  \bibinfo {author} {\bibfnamefont {T}~\bibnamefont {Kita}}, \bibinfo {author}
  {\bibfnamefont {Y}~\bibnamefont {Ohno}}, \bibinfo {author} {\bibfnamefont
  {H}~\bibnamefont {Ohno}}, \ and\ \bibinfo {author} {\bibfnamefont
  {M}~\bibnamefont {Kawasaki}},\ }\bibfield  {title} {\enquote {\bibinfo
  {title} {{Quantum Hall effect in polar oxide heterostructures}},}\ }\href
  {\doibase 10.1126/science.1137430} {\bibfield  {journal} {\bibinfo  {journal}
  {Science}\ }\textbf {\bibinfo {volume} {315}},\ \bibinfo {pages} {1388--1391}
  (\bibinfo {year} {2007})}\BibitemShut {NoStop}%
\bibitem [{\citenamefont {Miyake}\ \emph {et~al.}(2013)\citenamefont {Miyake},
  \citenamefont {Siviloglou}, \citenamefont {Kennedy}, \citenamefont {Burton},\
  and\ \citenamefont {Ketterle}}]{Miyake:2013jw}%
  \BibitemOpen
  \bibfield  {author} {\bibinfo {author} {\bibfnamefont {Hirokazu}\
  \bibnamefont {Miyake}}, \bibinfo {author} {\bibfnamefont {Georgios~A}\
  \bibnamefont {Siviloglou}}, \bibinfo {author} {\bibfnamefont {Colin~J}\
  \bibnamefont {Kennedy}}, \bibinfo {author} {\bibfnamefont {William~Cody}\
  \bibnamefont {Burton}}, \ and\ \bibinfo {author} {\bibfnamefont {Wolfgang}\
  \bibnamefont {Ketterle}},\ }\bibfield  {title} {\enquote {\bibinfo {title}
  {{Realizing the Harper Hamiltonian with Laser-Assisted Tunneling in Optical
  Lattices}},}\ }\href {\doibase 10.1103/PhysRevLett.111.185302} {\bibfield
  {journal} {\bibinfo  {journal} {Physical Review Letters}\ }\textbf {\bibinfo
  {volume} {111}},\ \bibinfo {pages} {185302} (\bibinfo {year}
  {2013})}\BibitemShut {NoStop}%
\bibitem [{\citenamefont {Aidelsburger}\ \emph {et~al.}(2013)\citenamefont
  {Aidelsburger}, \citenamefont {Atala}, \citenamefont {Lohse}, \citenamefont
  {Barreiro}, \citenamefont {Paredes},\ and\ \citenamefont
  {Bloch}}]{Aidelsburger:2013ew}%
  \BibitemOpen
  \bibfield  {author} {\bibinfo {author} {\bibfnamefont {M}~\bibnamefont
  {Aidelsburger}}, \bibinfo {author} {\bibfnamefont {M}~\bibnamefont {Atala}},
  \bibinfo {author} {\bibfnamefont {M}~\bibnamefont {Lohse}}, \bibinfo {author}
  {\bibfnamefont {J~T}\ \bibnamefont {Barreiro}}, \bibinfo {author}
  {\bibfnamefont {B}~\bibnamefont {Paredes}}, \ and\ \bibinfo {author}
  {\bibfnamefont {Immanuel}\ \bibnamefont {Bloch}},\ }\bibfield  {title}
  {\enquote {\bibinfo {title} {{Realization of the Hofstadter Hamiltonian with
  Ultracold Atoms in Optical Lattices}},}\ }\href {\doibase
  10.1103/PhysRevLett.111.185301} {\bibfield  {journal} {\bibinfo  {journal}
  {Physical Review Letters}\ }\textbf {\bibinfo {volume} {111}},\ \bibinfo
  {pages} {185301} (\bibinfo {year} {2013})}\BibitemShut {NoStop}%
\bibitem [{\citenamefont {Zhang}\ \emph {et~al.}(2005)\citenamefont {Zhang},
  \citenamefont {Tan}, \citenamefont {Stormer},\ and\ \citenamefont
  {Kim}}]{Zhang:2005gp}%
  \BibitemOpen
  \bibfield  {author} {\bibinfo {author} {\bibfnamefont {Yuanbo}\ \bibnamefont
  {Zhang}}, \bibinfo {author} {\bibfnamefont {Yan-Wen}\ \bibnamefont {Tan}},
  \bibinfo {author} {\bibfnamefont {Horst~L}\ \bibnamefont {Stormer}}, \ and\
  \bibinfo {author} {\bibfnamefont {Philip}\ \bibnamefont {Kim}},\ }\bibfield
  {title} {\enquote {\bibinfo {title} {{Experimental observation of the quantum
  Hall effect and Berry's phase in graphene}},}\ }\href {\doibase
  10.1038/nature04235} {\bibfield  {journal} {\bibinfo  {journal} {Nature}\
  }\textbf {\bibinfo {volume} {438}},\ \bibinfo {pages} {201--204} (\bibinfo
  {year} {2005})}\BibitemShut {NoStop}%
\bibitem [{\citenamefont {Novoselov}\ \emph {et~al.}(2005)\citenamefont
  {Novoselov}, \citenamefont {Geim}, \citenamefont {Morozov}, \citenamefont
  {Jiang}, \citenamefont {Katsnelson}, \citenamefont {Grigorieva},
  \citenamefont {Dubonos},\ and\ \citenamefont {Firsov}}]{Novoselov:2005es}%
  \BibitemOpen
  \bibfield  {author} {\bibinfo {author} {\bibfnamefont {K~S}\ \bibnamefont
  {Novoselov}}, \bibinfo {author} {\bibfnamefont {A~K}\ \bibnamefont {Geim}},
  \bibinfo {author} {\bibfnamefont {S~V}\ \bibnamefont {Morozov}}, \bibinfo
  {author} {\bibfnamefont {D}~\bibnamefont {Jiang}}, \bibinfo {author}
  {\bibfnamefont {M~I}\ \bibnamefont {Katsnelson}}, \bibinfo {author}
  {\bibfnamefont {I~V}\ \bibnamefont {Grigorieva}}, \bibinfo {author}
  {\bibfnamefont {S~V}\ \bibnamefont {Dubonos}}, \ and\ \bibinfo {author}
  {\bibfnamefont {A~A}\ \bibnamefont {Firsov}},\ }\bibfield  {title} {\enquote
  {\bibinfo {title} {{Two-dimensional gas of massless Dirac fermions in
  graphene}},}\ }\href {\doibase 10.1038/nature04233} {\bibfield  {journal}
  {\bibinfo  {journal} {Nature}\ }\textbf {\bibinfo {volume} {438}},\ \bibinfo
  {pages} {197--200} (\bibinfo {year} {2005})}\BibitemShut {NoStop}%
\bibitem [{\citenamefont {Xu}\ \emph {et~al.}(2014)\citenamefont {Xu},
  \citenamefont {Miotkowski}, \citenamefont {Liu}, \citenamefont {Tian},
  \citenamefont {Nam}, \citenamefont {Alidoust}, \citenamefont {Hu},
  \citenamefont {Shih}, \citenamefont {Hasan},\ and\ \citenamefont
  {Chen}}]{Xu:2014eh}%
  \BibitemOpen
  \bibfield  {author} {\bibinfo {author} {\bibfnamefont {Yang}\ \bibnamefont
  {Xu}}, \bibinfo {author} {\bibfnamefont {Ireneusz}\ \bibnamefont
  {Miotkowski}}, \bibinfo {author} {\bibfnamefont {Chang}\ \bibnamefont {Liu}},
  \bibinfo {author} {\bibfnamefont {Jifa}\ \bibnamefont {Tian}}, \bibinfo
  {author} {\bibfnamefont {Hyoungdo}\ \bibnamefont {Nam}}, \bibinfo {author}
  {\bibfnamefont {Nasser}\ \bibnamefont {Alidoust}}, \bibinfo {author}
  {\bibfnamefont {Jiuning}\ \bibnamefont {Hu}}, \bibinfo {author}
  {\bibfnamefont {Chih-Kang}\ \bibnamefont {Shih}}, \bibinfo {author}
  {\bibfnamefont {M~Zahid}\ \bibnamefont {Hasan}}, \ and\ \bibinfo {author}
  {\bibfnamefont {Yong~P}\ \bibnamefont {Chen}},\ }\bibfield  {title} {\enquote
  {\bibinfo {title} {{Observation of topological surface state quantum Hall
  effect in an intrinsic three-dimensional topological insulator}},}\ }\href
  {\doibase 10.1038/nphys3140} {\bibfield  {journal} {\bibinfo  {journal}
  {Nature Physics}\ }\textbf {\bibinfo {volume} {10}},\ \bibinfo {pages}
  {956--963} (\bibinfo {year} {2014})}\BibitemShut {NoStop}%
\bibitem [{\citenamefont {Berry}(1984)}]{Berry:1984ka}%
  \BibitemOpen
  \bibfield  {author} {\bibinfo {author} {\bibfnamefont {M~V}\ \bibnamefont
  {Berry}},\ }\bibfield  {title} {\enquote {\bibinfo {title} {{Quantal Phase
  Factors Accompanying Adiabatic Changes}},}\ }\href {\doibase
  10.1098/rspa.1984.0023} {\bibfield  {journal} {\bibinfo  {journal}
  {Proceedings of the Royal Society A: Mathematical, Physical and Engineering
  Sciences}\ }\textbf {\bibinfo {volume} {392}},\ \bibinfo {pages} {45--57}
  (\bibinfo {year} {1984})}\BibitemShut {NoStop}%
\bibitem [{\citenamefont {Parameswaran}\ \emph {et~al.}(2012)\citenamefont
  {Parameswaran}, \citenamefont {Roy},\ and\ \citenamefont
  {Sondhi}}]{Parameswaran:2012uk}%
  \BibitemOpen
  \bibfield  {author} {\bibinfo {author} {\bibfnamefont {S~A}\ \bibnamefont
  {Parameswaran}}, \bibinfo {author} {\bibfnamefont {R}~\bibnamefont {Roy}}, \
  and\ \bibinfo {author} {\bibfnamefont {Shivaji~L}\ \bibnamefont {Sondhi}},\
  }\bibfield  {title} {\enquote {\bibinfo {title} {{Fractional Chern insulators
  and the W$\infty$ algebra}},}\ }\href
  {http://prb.aps.org/abstract/PRB/v85/i24/e241308} {\bibfield  {journal}
  {\bibinfo  {journal} {Physical Review B}\ }\textbf {\bibinfo {volume} {85}},\
  \bibinfo {pages} {241308} (\bibinfo {year} {2012})}\BibitemShut {NoStop}%
\bibitem [{\citenamefont {Roy}(2014)}]{Roy:2014dl}%
  \BibitemOpen
  \bibfield  {author} {\bibinfo {author} {\bibfnamefont {Rahul}\ \bibnamefont
  {Roy}},\ }\bibfield  {title} {\enquote {\bibinfo {title} {{Band geometry of
  fractional topological insulators}},}\ }\href {\doibase
  10.1103/PhysRevB.90.165139} {\bibfield  {journal} {\bibinfo  {journal}
  {Physical Review B}\ }\textbf {\bibinfo {volume} {90}},\ \bibinfo {pages}
  {165139} (\bibinfo {year} {2014})}\BibitemShut {NoStop}%
\bibitem [{\citenamefont {Bergholtz}\ and\ \citenamefont
  {Liu}(2013)}]{Bergholtz:2013ey}%
  \BibitemOpen
  \bibfield  {author} {\bibinfo {author} {\bibfnamefont {Emil~J}\ \bibnamefont
  {Bergholtz}}\ and\ \bibinfo {author} {\bibfnamefont {Zhao}\ \bibnamefont
  {Liu}},\ }\bibfield  {title} {\enquote {\bibinfo {title} {{Topological Flat
  Band Models and Fractional Chern Insulators}},}\ }\href {\doibase
  10.1142/S021797921330017X} {\bibfield  {journal} {\bibinfo  {journal}
  {International Journal of Modern Physics B}\ }\textbf {\bibinfo {volume}
  {27}},\ \bibinfo {pages} {1330017--1--43} (\bibinfo {year}
  {2013})}\BibitemShut {NoStop}%
\bibitem [{\citenamefont {Parameswaran}\ \emph {et~al.}(2013)\citenamefont
  {Parameswaran}, \citenamefont {Roy},\ and\ \citenamefont
  {Sondhi}}]{Parameswaran:2013pca}%
  \BibitemOpen
  \bibfield  {author} {\bibinfo {author} {\bibfnamefont {Siddharth~A}\
  \bibnamefont {Parameswaran}}, \bibinfo {author} {\bibfnamefont {Rahul}\
  \bibnamefont {Roy}}, \ and\ \bibinfo {author} {\bibfnamefont {Shivaji~L}\
  \bibnamefont {Sondhi}},\ }\bibfield  {title} {\enquote {\bibinfo {title}
  {{Comptes Rendus Physique}},}\ }\href {\doibase 10.1016/j.crhy.2013.04.003}
  {\bibfield  {journal} {\bibinfo  {journal} {Comptes Rendus Physique}\
  }\textbf {\bibinfo {volume} {14}},\ \bibinfo {pages} {816--839} (\bibinfo
  {year} {2013})}\BibitemShut {NoStop}%
\bibitem [{\citenamefont {Harper}\ \emph {et~al.}(2014)\citenamefont {Harper},
  \citenamefont {Simon},\ and\ \citenamefont {Roy}}]{Harper:2014fq}%
  \BibitemOpen
  \bibfield  {author} {\bibinfo {author} {\bibfnamefont {Fenner}\ \bibnamefont
  {Harper}}, \bibinfo {author} {\bibfnamefont {Steven~H}\ \bibnamefont
  {Simon}}, \ and\ \bibinfo {author} {\bibfnamefont {Rahul}\ \bibnamefont
  {Roy}},\ }\bibfield  {title} {\enquote {\bibinfo {title} {{Perturbative
  approach to flat Chern bands in the Hofstadter model}},}\ }\href {\doibase
  10.1103/PhysRevB.90.075104} {\bibfield  {journal} {\bibinfo  {journal}
  {Physical Review B}\ }\textbf {\bibinfo {volume} {90}},\ \bibinfo {pages}
  {075104} (\bibinfo {year} {2014})}\BibitemShut {NoStop}%
\bibitem [{\citenamefont {Bauer}\ \emph {et~al.}(2016)\citenamefont {Bauer},
  \citenamefont {Jackson},\ and\ \citenamefont {Roy}}]{Bauer:2016ju}%
  \BibitemOpen
  \bibfield  {author} {\bibinfo {author} {\bibfnamefont {David}\ \bibnamefont
  {Bauer}}, \bibinfo {author} {\bibfnamefont {T~S}\ \bibnamefont {Jackson}}, \
  and\ \bibinfo {author} {\bibfnamefont {Rahul}\ \bibnamefont {Roy}},\
  }\bibfield  {title} {\enquote {\bibinfo {title} {{Quantum geometry and
  stability of the fractional quantum Hall effect in the Hofstadter model}},}\
  }\href {\doibase 10.1103/PhysRevB.93.235133} {\bibfield  {journal} {\bibinfo
  {journal} {Physical Review B}\ }\textbf {\bibinfo {volume} {93}},\ \bibinfo
  {pages} {235133--11} (\bibinfo {year} {2016})}\BibitemShut {NoStop}%
\bibitem [{\citenamefont {Xiao}\ \emph {et~al.}(2010)\citenamefont {Xiao},
  \citenamefont {Chang},\ and\ \citenamefont {Niu}}]{Xiao:2010kw}%
  \BibitemOpen
  \bibfield  {author} {\bibinfo {author} {\bibfnamefont {Di}~\bibnamefont
  {Xiao}}, \bibinfo {author} {\bibfnamefont {Ming-Che}\ \bibnamefont {Chang}},
  \ and\ \bibinfo {author} {\bibfnamefont {Qian}\ \bibnamefont {Niu}},\
  }\bibfield  {title} {\enquote {\bibinfo {title} {{Berry phase effects on
  electronic properties}},}\ }\href {\doibase 10.1103/RevModPhys.82.1959}
  {\bibfield  {journal} {\bibinfo  {journal} {Reviews of Modern Physics}\
  }\textbf {\bibinfo {volume} {82}},\ \bibinfo {pages} {1959--2007} (\bibinfo
  {year} {2010})}\BibitemShut {NoStop}%
\bibitem [{\citenamefont {Jackson}\ \emph {et~al.}(2015)\citenamefont
  {Jackson}, \citenamefont {Moller},\ and\ \citenamefont
  {Roy}}]{Jackson:2015fv}%
  \BibitemOpen
  \bibfield  {author} {\bibinfo {author} {\bibfnamefont {T~S}\ \bibnamefont
  {Jackson}}, \bibinfo {author} {\bibfnamefont {Gunnar}\ \bibnamefont
  {Moller}}, \ and\ \bibinfo {author} {\bibfnamefont {Rahul}\ \bibnamefont
  {Roy}},\ }\bibfield  {title} {\enquote {\bibinfo {title} {{Geometric
  stability of topological lattice phases}},}\ }\href {\doibase
  10.1038/ncomms9629} {\bibfield  {journal} {\bibinfo  {journal} {Nature
  Communications}\ ,\ \bibinfo {pages} {1--11}} (\bibinfo {year}
  {2015})}\BibitemShut {NoStop}%
\bibitem [{\citenamefont {Haldane}(1988)}]{Haldane:1988uf}%
  \BibitemOpen
  \bibfield  {author} {\bibinfo {author} {\bibfnamefont {FDM}\ \bibnamefont
  {Haldane}},\ }\bibfield  {title} {\enquote {\bibinfo {title} {{Model for a
  quantum Hall effect without Landau levels: Condensed-matter realization of
  the" parity anomaly"}},}\ }\href
  {http://link.aps.org/doi/10.1103/PhysRevLett.61.2015} {\bibfield  {journal}
  {\bibinfo  {journal} {Physical Review Letters}\ }\textbf {\bibinfo {volume}
  {61}},\ \bibinfo {pages} {2015--2018} (\bibinfo {year} {1988})}\BibitemShut
  {NoStop}%
\bibitem [{\citenamefont {Neupert}\ \emph {et~al.}(2011)\citenamefont
  {Neupert}, \citenamefont {Santos}, \citenamefont {Chamon},\ and\
  \citenamefont {Mudry}}]{Neupert:2011db}%
  \BibitemOpen
  \bibfield  {author} {\bibinfo {author} {\bibfnamefont {Titus}\ \bibnamefont
  {Neupert}}, \bibinfo {author} {\bibfnamefont {Luiz}\ \bibnamefont {Santos}},
  \bibinfo {author} {\bibfnamefont {Claudio}\ \bibnamefont {Chamon}}, \ and\
  \bibinfo {author} {\bibfnamefont {Christopher}\ \bibnamefont {Mudry}},\
  }\bibfield  {title} {\enquote {\bibinfo {title} {{Fractional Quantum Hall
  States at Zero Magnetic Field}},}\ }\href {\doibase
  10.1103/PhysRevLett.106.236804} {\bibfield  {journal} {\bibinfo  {journal}
  {Physical Review Letters}\ }\textbf {\bibinfo {volume} {106}},\ \bibinfo
  {pages} {236804} (\bibinfo {year} {2011})}\BibitemShut {NoStop}%
\bibitem [{\citenamefont {Sun}\ \emph {et~al.}(2011)\citenamefont {Sun},
  \citenamefont {Gu}, \citenamefont {Katsura},\ and\ \citenamefont
  {Das~Sarma}}]{Sun:2011dk}%
  \BibitemOpen
  \bibfield  {author} {\bibinfo {author} {\bibfnamefont {Kai}\ \bibnamefont
  {Sun}}, \bibinfo {author} {\bibfnamefont {Zhengcheng}\ \bibnamefont {Gu}},
  \bibinfo {author} {\bibfnamefont {Hosho}\ \bibnamefont {Katsura}}, \ and\
  \bibinfo {author} {\bibfnamefont {S}~\bibnamefont {Das~Sarma}},\ }\bibfield
  {title} {\enquote {\bibinfo {title} {{Nearly Flatbands with Nontrivial
  Topology}},}\ }\href {\doibase 10.1103/PhysRevLett.106.236803} {\bibfield
  {journal} {\bibinfo  {journal} {Physical Review Letters}\ }\textbf {\bibinfo
  {volume} {106}},\ \bibinfo {pages} {236803} (\bibinfo {year}
  {2011})}\BibitemShut {NoStop}%
\bibitem [{\citenamefont {Sheng}\ \emph {et~al.}(2011)\citenamefont {Sheng},
  \citenamefont {Gu}, \citenamefont {Sun},\ and\ \citenamefont
  {Sheng}}]{Sheng:2011iv}%
  \BibitemOpen
  \bibfield  {author} {\bibinfo {author} {\bibfnamefont {D~N}\ \bibnamefont
  {Sheng}}, \bibinfo {author} {\bibfnamefont {Zheng-Cheng}\ \bibnamefont {Gu}},
  \bibinfo {author} {\bibfnamefont {Kai}\ \bibnamefont {Sun}}, \ and\ \bibinfo
  {author} {\bibfnamefont {L}~\bibnamefont {Sheng}},\ }\bibfield  {title}
  {\enquote {\bibinfo {title} {{Fractional quantum Hall effect in the absence
  of Landau levels}},}\ }\href {\doibase 10.1038/ncomms1380} {\bibfield
  {journal} {\bibinfo  {journal} {Nature Communications}\ }\textbf {\bibinfo
  {volume} {2}},\ \bibinfo {pages} {389--385} (\bibinfo {year}
  {2011})}\BibitemShut {NoStop}%
\bibitem [{\citenamefont {Chang}\ \emph {et~al.}(2013)\citenamefont {Chang},
  \citenamefont {Zhang}, \citenamefont {Feng}, \citenamefont {Shen},
  \citenamefont {Zhang}, \citenamefont {Guo}, \citenamefont {Li}, \citenamefont
  {Ou}, \citenamefont {Wei}, \citenamefont {Wang}, \citenamefont {Ji},
  \citenamefont {Feng}, \citenamefont {Ji}, \citenamefont {Chen}, \citenamefont
  {Jia}, \citenamefont {Dai}, \citenamefont {Fang}, \citenamefont {Zhang},
  \citenamefont {He}, \citenamefont {Wang}, \citenamefont {Lu}, \citenamefont
  {Ma},\ and\ \citenamefont {Xue}}]{Chang:2013dd}%
  \BibitemOpen
  \bibfield  {author} {\bibinfo {author} {\bibfnamefont {Cui-Zu}\ \bibnamefont
  {Chang}}, \bibinfo {author} {\bibfnamefont {Jinsong}\ \bibnamefont {Zhang}},
  \bibinfo {author} {\bibfnamefont {Xiao}\ \bibnamefont {Feng}}, \bibinfo
  {author} {\bibfnamefont {Jie}\ \bibnamefont {Shen}}, \bibinfo {author}
  {\bibfnamefont {Zuocheng}\ \bibnamefont {Zhang}}, \bibinfo {author}
  {\bibfnamefont {Minghua}\ \bibnamefont {Guo}}, \bibinfo {author}
  {\bibfnamefont {Kang}\ \bibnamefont {Li}}, \bibinfo {author} {\bibfnamefont
  {Yunbo}\ \bibnamefont {Ou}}, \bibinfo {author} {\bibfnamefont {Pang}\
  \bibnamefont {Wei}}, \bibinfo {author} {\bibfnamefont {Li-Li}\ \bibnamefont
  {Wang}}, \bibinfo {author} {\bibfnamefont {Zhong-Qing}\ \bibnamefont {Ji}},
  \bibinfo {author} {\bibfnamefont {Yang}\ \bibnamefont {Feng}}, \bibinfo
  {author} {\bibfnamefont {Shuaihua}\ \bibnamefont {Ji}}, \bibinfo {author}
  {\bibfnamefont {Xi}~\bibnamefont {Chen}}, \bibinfo {author} {\bibfnamefont
  {Jinfeng}\ \bibnamefont {Jia}}, \bibinfo {author} {\bibfnamefont
  {Xi}~\bibnamefont {Dai}}, \bibinfo {author} {\bibfnamefont {Zhong}\
  \bibnamefont {Fang}}, \bibinfo {author} {\bibfnamefont {Shou-Cheng}\
  \bibnamefont {Zhang}}, \bibinfo {author} {\bibfnamefont {Ke}~\bibnamefont
  {He}}, \bibinfo {author} {\bibfnamefont {Yayu}\ \bibnamefont {Wang}},
  \bibinfo {author} {\bibfnamefont {Li}~\bibnamefont {Lu}}, \bibinfo {author}
  {\bibfnamefont {Xu-Cun}\ \bibnamefont {Ma}}, \ and\ \bibinfo {author}
  {\bibfnamefont {Qi-Kun}\ \bibnamefont {Xue}},\ }\bibfield  {title} {\enquote
  {\bibinfo {title} {{Experimental Observation of the Quantum Anomalous Hall
  Effect in a Magnetic Topological Insulator}},}\ }\href {\doibase
  10.1126/science.1234414} {\bibfield  {journal} {\bibinfo  {journal}
  {Science}\ }\textbf {\bibinfo {volume} {340}},\ \bibinfo {pages} {167--170}
  (\bibinfo {year} {2013})}\BibitemShut {NoStop}%
\bibitem [{\citenamefont {He}\ \emph {et~al.}(2017)\citenamefont {He},
  \citenamefont {Li}, \citenamefont {Lyu},\ and\ \citenamefont
  {Nachtigall}}]{He:2017ib}%
  \BibitemOpen
  \bibfield  {author} {\bibinfo {author} {\bibfnamefont {Junjie}\ \bibnamefont
  {He}}, \bibinfo {author} {\bibfnamefont {Xiao}\ \bibnamefont {Li}}, \bibinfo
  {author} {\bibfnamefont {Pengbo}\ \bibnamefont {Lyu}}, \ and\ \bibinfo
  {author} {\bibfnamefont {Petr}\ \bibnamefont {Nachtigall}},\ }\bibfield
  {title} {\enquote {\bibinfo {title} {{Near-Room-Temperature Chern Insulator
  and Dirac Spin-Gapless Semiconductor: Nickle Chloride Monolayer}},}\ }\href
  {\doibase 10.1039/C6NR08522A} {\bibfield  {journal} {\bibinfo  {journal}
  {Nanoscale}\ ,\ \bibinfo {pages} {1--19}} (\bibinfo {year}
  {2017})}\BibitemShut {NoStop}%
\bibitem [{\citenamefont {Dalibard}\ \emph {et~al.}(2011)\citenamefont
  {Dalibard}, \citenamefont {Gerbier}, \citenamefont {Juzeli{\=u}nas},\ and\
  \citenamefont {{\"O}hberg}}]{Dalibard:2011gg}%
  \BibitemOpen
  \bibfield  {author} {\bibinfo {author} {\bibfnamefont {Jean}\ \bibnamefont
  {Dalibard}}, \bibinfo {author} {\bibfnamefont {Fabrice}\ \bibnamefont
  {Gerbier}}, \bibinfo {author} {\bibfnamefont {Gediminas}\ \bibnamefont
  {Juzeli{\=u}nas}}, \ and\ \bibinfo {author} {\bibfnamefont {Patrik}\
  \bibnamefont {{\"O}hberg}},\ }\bibfield  {title} {\enquote {\bibinfo {title}
  {{Colloquium: Artificial gauge potentials for neutral atoms}},}\ }\href
  {\doibase 10.1103/RevModPhys.83.1523} {\bibfield  {journal} {\bibinfo
  {journal} {Reviews of Modern Physics}\ }\textbf {\bibinfo {volume} {83}},\
  \bibinfo {pages} {1523--1543} (\bibinfo {year} {2011})}\BibitemShut {NoStop}%
\bibitem [{\citenamefont {Ponomarenko}\ \emph {et~al.}(2013)\citenamefont
  {Ponomarenko}, \citenamefont {Gorbachev}, \citenamefont {Yu}, \citenamefont
  {Elias}, \citenamefont {Jalil}, \citenamefont {Patel}, \citenamefont
  {Mishchenko}, \citenamefont {Mayorov}, \citenamefont {Woods}, \citenamefont
  {Wallbank}, \citenamefont {Mucha-Kruczynski}, \citenamefont {Piot},
  \citenamefont {Potemski}, \citenamefont {Grigorieva}, \citenamefont
  {Novoselov}, \citenamefont {Guinea}, \citenamefont {Fal{\textquoteright}ko},\
  and\ \citenamefont {Geim}}]{Ponomarenko:2014hl}%
  \BibitemOpen
  \bibfield  {author} {\bibinfo {author} {\bibfnamefont {L~A}\ \bibnamefont
  {Ponomarenko}}, \bibinfo {author} {\bibfnamefont {R~V}\ \bibnamefont
  {Gorbachev}}, \bibinfo {author} {\bibfnamefont {G~L}\ \bibnamefont {Yu}},
  \bibinfo {author} {\bibfnamefont {D~C}\ \bibnamefont {Elias}}, \bibinfo
  {author} {\bibfnamefont {R}~\bibnamefont {Jalil}}, \bibinfo {author}
  {\bibfnamefont {A~A}\ \bibnamefont {Patel}}, \bibinfo {author} {\bibfnamefont
  {A}~\bibnamefont {Mishchenko}}, \bibinfo {author} {\bibfnamefont {A~S}\
  \bibnamefont {Mayorov}}, \bibinfo {author} {\bibfnamefont {C~R}\ \bibnamefont
  {Woods}}, \bibinfo {author} {\bibfnamefont {J~R}\ \bibnamefont {Wallbank}},
  \bibinfo {author} {\bibfnamefont {M}~\bibnamefont {Mucha-Kruczynski}},
  \bibinfo {author} {\bibfnamefont {B~A}\ \bibnamefont {Piot}}, \bibinfo
  {author} {\bibfnamefont {M}~\bibnamefont {Potemski}}, \bibinfo {author}
  {\bibfnamefont {I~V}\ \bibnamefont {Grigorieva}}, \bibinfo {author}
  {\bibfnamefont {K~S}\ \bibnamefont {Novoselov}}, \bibinfo {author}
  {\bibfnamefont {F}~\bibnamefont {Guinea}}, \bibinfo {author} {\bibfnamefont
  {V~I}\ \bibnamefont {Fal{\textquoteright}ko}}, \ and\ \bibinfo {author}
  {\bibfnamefont {A~K}\ \bibnamefont {Geim}},\ }\bibfield  {title} {\enquote
  {\bibinfo {title} {{Cloning of Dirac fermions in graphene superlattices}},}\
  }\href {\doibase 10.1038/nature12187} {\bibfield  {journal} {\bibinfo
  {journal} {Nature}\ }\textbf {\bibinfo {volume} {497}},\ \bibinfo {pages}
  {594--597} (\bibinfo {year} {2013})}\BibitemShut {NoStop}%
\bibitem [{\citenamefont {Dean}\ \emph {et~al.}(2013)\citenamefont {Dean},
  \citenamefont {Wang}, \citenamefont {Maher}, \citenamefont {Forsythe},
  \citenamefont {Ghahari}, \citenamefont {Gao}, \citenamefont {Katoch},
  \citenamefont {Ishigami}, \citenamefont {Moon}, \citenamefont {Koshino},
  \citenamefont {Taniguchi}, \citenamefont {Watanabe}, \citenamefont {Shepard},
  \citenamefont {Hone},\ and\ \citenamefont {Kim}}]{Dean:2014bv}%
  \BibitemOpen
  \bibfield  {author} {\bibinfo {author} {\bibfnamefont {C~R}\ \bibnamefont
  {Dean}}, \bibinfo {author} {\bibfnamefont {L}~\bibnamefont {Wang}}, \bibinfo
  {author} {\bibfnamefont {P}~\bibnamefont {Maher}}, \bibinfo {author}
  {\bibfnamefont {C}~\bibnamefont {Forsythe}}, \bibinfo {author} {\bibfnamefont
  {F}~\bibnamefont {Ghahari}}, \bibinfo {author} {\bibfnamefont
  {Y}~\bibnamefont {Gao}}, \bibinfo {author} {\bibfnamefont {J}~\bibnamefont
  {Katoch}}, \bibinfo {author} {\bibfnamefont {M}~\bibnamefont {Ishigami}},
  \bibinfo {author} {\bibfnamefont {P}~\bibnamefont {Moon}}, \bibinfo {author}
  {\bibfnamefont {M}~\bibnamefont {Koshino}}, \bibinfo {author} {\bibfnamefont
  {T}~\bibnamefont {Taniguchi}}, \bibinfo {author} {\bibfnamefont
  {K}~\bibnamefont {Watanabe}}, \bibinfo {author} {\bibfnamefont {K~L}\
  \bibnamefont {Shepard}}, \bibinfo {author} {\bibfnamefont {J}~\bibnamefont
  {Hone}}, \ and\ \bibinfo {author} {\bibfnamefont {P}~\bibnamefont {Kim}},\
  }\bibfield  {title} {\enquote {\bibinfo {title} {{Hofstadter's butterfly and
  the fractal quantum Hall effect in moire ́ superlattices}},}\ }\href
  {\doibase 10.1038/nature12186} {\bibfield  {journal} {\bibinfo  {journal}
  {Nature}\ }\textbf {\bibinfo {volume} {497}},\ \bibinfo {pages} {598--602}
  (\bibinfo {year} {2013})}\BibitemShut {NoStop}%
\bibitem [{\citenamefont {Harper}(1955)}]{Harper:1955uu}%
  \BibitemOpen
  \bibfield  {author} {\bibinfo {author} {\bibfnamefont {P~G}\ \bibnamefont
  {Harper}},\ }\bibfield  {title} {\enquote {\bibinfo {title} {{The general
  motion of conduction electrons in a uniform magnetic field, with application
  to the diamagnetism of metals}},}\ }\href
  {http://iopscience.iop.org/0370-1298/68/10/305} {\bibfield  {journal}
  {\bibinfo  {journal} {Proceedings of the Royal Society A: Mathematical,
  Physical and Engineering Sciences}\ }\textbf {\bibinfo {volume} {68}},\
  \bibinfo {pages} {879} (\bibinfo {year} {1955})}\BibitemShut {NoStop}%
\bibitem [{\citenamefont {Hofstadter}(1976)}]{Hofstadter:1976wt}%
  \BibitemOpen
  \bibfield  {author} {\bibinfo {author} {\bibfnamefont {D~R}\ \bibnamefont
  {Hofstadter}},\ }\bibfield  {title} {\enquote {\bibinfo {title} {{Energy
  levels and wave functions of Bloch electrons in rational and irrational
  magnetic fields}},}\ }\href {http://prb.aps.org/abstract/PRB/v14/i6/p2239_1}
  {\bibfield  {journal} {\bibinfo  {journal} {Physical Review B}\ }\textbf
  {\bibinfo {volume} {14}},\ \bibinfo {pages} {2239} (\bibinfo {year}
  {1976})}\BibitemShut {NoStop}%
\bibitem [{\citenamefont {Landau}\ and\ \citenamefont
  {Lifshit︠z}(1977)}]{landau1977quantum}%
  \BibitemOpen
  \bibfield  {author} {\bibinfo {author} {\bibfnamefont {L~D}\ \bibnamefont
  {Landau}}\ and\ \bibinfo {author} {\bibfnamefont {E~M}\ \bibnamefont
  {Lifshit︠z}},\ }\href {https://books.google.co.uk/books?id=J9ui6KwC4mMC}
  {\emph {\bibinfo {title} {{Quantum Mechanics: Non-relativistic Theory}}}},\
  Butterworth Heinemann\ (\bibinfo  {publisher} {Butterworth-Heinemann},\
  \bibinfo {address} {Oxford},\ \bibinfo {year} {1977})\BibitemShut {NoStop}%
\bibitem [{\citenamefont {Peierls}(1933)}]{Peierls:1933tv}%
  \BibitemOpen
  \bibfield  {author} {\bibinfo {author} {\bibfnamefont {Rudolph}\ \bibnamefont
  {Peierls}},\ }\bibfield  {title} {\enquote {\bibinfo {title} {{Zur theorie
  des diamagnetismus von leitungselektronen}},}\ }\href
  {http://link.springer.com/article/10.1007/BF01342591} {\bibfield  {journal}
  {\bibinfo  {journal} {Zeitschrift f{\"u}r Physik}\ }\textbf {\bibinfo
  {volume} {80}},\ \bibinfo {pages} {763--791} (\bibinfo {year}
  {1933})}\BibitemShut {NoStop}%
\bibitem [{\citenamefont {Luttinger}(1951)}]{Luttinger:1951bv}%
  \BibitemOpen
  \bibfield  {author} {\bibinfo {author} {\bibfnamefont {J~M}\ \bibnamefont
  {Luttinger}},\ }\bibfield  {title} {\enquote {\bibinfo {title} {{The Effect
  of a Magnetic Field on Electrons in a Periodic Potential}},}\ }\href
  {\doibase 10.1103/PhysRev.84.814} {\bibfield  {journal} {\bibinfo  {journal}
  {Physical Review}\ }\textbf {\bibinfo {volume} {84}},\ \bibinfo {pages}
  {814--817} (\bibinfo {year} {1951})}\BibitemShut {NoStop}%
\bibitem [{\citenamefont {Kohn}(1959)}]{Kohn:1959zza}%
  \BibitemOpen
  \bibfield  {author} {\bibinfo {author} {\bibfnamefont {Walter}\ \bibnamefont
  {Kohn}},\ }\bibfield  {title} {\enquote {\bibinfo {title} {{Theory of Bloch
  Electrons in a Magnetic Field: The Effective Hamiltonian}},}\ }\href
  {\doibase 10.1103/PhysRev.115.1460} {\bibfield  {journal} {\bibinfo
  {journal} {Physical Review}\ }\textbf {\bibinfo {volume} {115}},\ \bibinfo
  {pages} {1460--1478} (\bibinfo {year} {1959})}\BibitemShut {NoStop}%
\bibitem [{\citenamefont {Blount}(1962)}]{Blount:1962jn}%
  \BibitemOpen
  \bibfield  {author} {\bibinfo {author} {\bibfnamefont {E~I}\ \bibnamefont
  {Blount}},\ }\bibfield  {title} {\enquote {\bibinfo {title} {{Bloch Electrons
  in a Magnetic Field}},}\ }\href {\doibase 10.1103/physrev.126.1636}
  {\bibfield  {journal} {\bibinfo  {journal} {Physical Review}\ }\textbf
  {\bibinfo {volume} {126}},\ \bibinfo {pages} {1636--1653} (\bibinfo {year}
  {1962})}\BibitemShut {NoStop}%
\bibitem [{\citenamefont {Wilkinson}(1987)}]{Wilkinson:1987tt}%
  \BibitemOpen
  \bibfield  {author} {\bibinfo {author} {\bibfnamefont {M}~\bibnamefont
  {Wilkinson}},\ }\bibfield  {title} {\enquote {\bibinfo {title} {{An Exact
  Renormalization-Group for Bloch Electrons in a Magnetic-Field}},}\ }\href
  {http://iopscience.iop.org/0305-4470/20/13/035} {\bibfield  {journal}
  {\bibinfo  {journal} {Journal of Physics A: Mathematical and General}\
  }\textbf {\bibinfo {volume} {20}},\ \bibinfo {pages} {4337--4354} (\bibinfo
  {year} {1987})}\BibitemShut {NoStop}%
\bibitem [{Note1()}]{Note1}%
  \BibitemOpen
  \bibinfo {note} {We leave a discussion of systems where the leading quadratic
  term vanishes to future work.}\BibitemShut {Stop}%
\bibitem [{\citenamefont {Mahan}(2000)}]{Mahan:2000wd}%
  \BibitemOpen
  \bibfield  {author} {\bibinfo {author} {\bibfnamefont {Gerald~D}\
  \bibnamefont {Mahan}},\ }\href
  {http://books.google.co.uk/books?id=xzSgZ4-yyMEC&printsec=frontcover&dq=many+particle+physics+mahan&hl=&cd=1&source=gbs_api}
  {\emph {\bibinfo {title} {{Many Particle Physics}}}}\ (\bibinfo  {publisher}
  {Plenum Publishing Corporation},\ \bibinfo {year} {2000})\BibitemShut
  {NoStop}%
\bibitem [{\citenamefont {Bradlyn}(2015)}]{Bradlyn:2015vw}%
  \BibitemOpen
  \bibfield  {author} {\bibinfo {author} {\bibfnamefont {B~J}\ \bibnamefont
  {Bradlyn}},\ }\emph {\bibinfo {title} {{Linear response and Berry curvature
  in two-dimensional topological phases}}},\ \href
  {http://adsabs.harvard.edu/abs/2015PhDT.......228B} {Ph.D. thesis} (\bibinfo
  {year} {2015})\BibitemShut {NoStop}%
\bibitem [{Note2()}]{Note2}%
  \BibitemOpen
  \bibinfo {note} {See Appendix~A of Ref.~\cite {Chen:1989xs} for an example of
  the latter approach in a slightly different context.}\BibitemShut {Stop}%
\bibitem [{\citenamefont {Chen}\ \emph {et~al.}(1989)\citenamefont {Chen},
  \citenamefont {Wilczek}, \citenamefont {Witten},\ and\ \citenamefont
  {Halperin}}]{Chen:1989xs}%
  \BibitemOpen
  \bibfield  {author} {\bibinfo {author} {\bibfnamefont {Yi-Hong}\ \bibnamefont
  {Chen}}, \bibinfo {author} {\bibfnamefont {Frank}\ \bibnamefont {Wilczek}},
  \bibinfo {author} {\bibfnamefont {Edward}\ \bibnamefont {Witten}}, \ and\
  \bibinfo {author} {\bibfnamefont {Bertrand~I}\ \bibnamefont {Halperin}},\
  }\bibfield  {title} {\enquote {\bibinfo {title} {{On Anyon
  Superconductivity}},}\ }\href {\doibase 10.1142/S0217979289000725} {\bibfield
   {journal} {\bibinfo  {journal} {Int. J. Mod. Phys.}\ }\textbf {\bibinfo
  {volume} {B3}},\ \bibinfo {pages} {1001--1067} (\bibinfo {year}
  {1989})}\BibitemShut {NoStop}%
\end{thebibliography}
\end{document}